\documentclass[aps,twocolumn,letterpaper,prd]{revtex4}

\usepackage{latexsym}
\usepackage{epsfig}
\usepackage{scrextend}
 \usepackage[french]{babel}

\usepackage{hyperref}
\usepackage[hyphenbreaks]{breakurl}

\usepackage{latexsym}
\usepackage{epsfig}
\usepackage{scrextend}
\usepackage{lipsum}

\newcommand{\footremember}[2]{%
    \footnote{#2}
    \newcounter{#1}
    \setcounter{#1}{\value{footnote}}%
}

\def\be{\begin{equation}}
\def\ee{\end{equation}}
\def\bea{\begin{eqnarray}}
\def\eea{\end{eqnarray}}
\def\ba{\begin{array}} 
\def\ea{\end{array}}
\def\bc{\begin{center}}
\def\ec{\end{center}}

\def\ghost#1{}

\def\simge{\mathrel{%
   \rlap{\raise 0.511ex \hbox{$>$}}{\lower 0.511ex \hbox{$\sim$}}}}
\def\simle{\mathrel{
   \rlap{\raise 0.511ex \hbox{$<$}}{\lower 0.511ex \hbox{$\sim$}}}}

\def\dis{\displaystyle}

\begin{document}

\renewcommand{\eprint}[1]{\href{https://arxiv.org/abs/#1}{arXiv:#1}}

\title{\boldmath  The $\,U$ boson, \,interpolating between \vspace{2mm}a generalized dark photon
\vspace{2mm}
\hbox{or dark $Z$, an axial boson and an axionlike particle}
\vspace{2MM}
}

\author{{\sc P}{ierre} {\sc Fayet}
\vspace{3mm} \\ \small }

\affiliation{Laboratoire de physique de l'\'Ecole normale sup\'erieure\vspace{.5mm}
\hspace*{-.8mm}\footremember{a}{\vspace*{0mm}  \hspace{-.8mm} \em 
ENS, Universit\'e PSL, CNRS, Sorbonne Universit\'e, Universit\'e de Paris, F-75005 Paris, France
}
\\ \hspace*{4mm}24 rue Lhomond, 75231 Paris cedex 05, France 
\vspace{1.5mm}\\
\hbox{and \,Centre de physique th\'eorique, \'Ecole polytechnique, 91128 Palaiseau cedex, France}
\vspace{2mm}\\
\vspace{1mm}}

\date{October 9, 2020}

\begin{abstract}

\vspace{1mm}

A light boson $U$ from an extra $U(1)$ interpolates between a generalized dark photon coupled to $\,Q,\,B$ and $L_i$ (or $B-L$), 
plus possibly dark matter, a dark $Z$ coupled to the $Z$ current, and one axially coupled to quarks and leptons. 
 We iden\-tify the corresponding $\,U(1)_F$ symmetries,  with
$F= \gamma_Y Y+\gamma_B B\,+\gamma_{L_i}L_i+\gamma_A F_A+\gamma_{F'}F'\!+\gamma_d \,F_d$, 
$F_d$ acting in a dark sector and $F'\!$ on possible semi-inert BEH doublets uncoupled to quarks and leptons.
The $U$ current is obtained from the $U(1)_F $ and $Z$ currents, with a mixing determined by the 
spin-0 BEH fields.
 The charge $Q_U$ of chiral quarks and leptons is a combination of $\,Q,\,B,\,L_i$ and $T_{3L}$ with the axial~$F_A$. It involves in general isovector and isoscalar axial terms, in the presence of two BEH doublets.
 
 \vspace{1mm}
 
A longitudinal $U$ with axial couplings has enhanced interactions, and behaves much as an axionlike particle.
 \vspace{-.7mm}
 Its axial couplings $g_A$, usually restricted to $\,\simle 2\times 10^{-7}\,m_U$(MeV),
 lead to effective pseudoscalar ones 
$\,g_P=  g_A\times 2m_{q,l}/m_U=$ $2^{1/4}\,G_F^{1/2}\,m_{q,l} \,A_\pm \,$. $A_\pm$ is proportional to an invisibility parameter $r=\cos\theta_A$ induced by a singlet v.e.v., possibly large and allowing the $U$ to be very weakly interacting.
 
 \vspace{1mm}
 This allows for a very small gauge coupling, expressed with two doublets and a singlet as $\,g"/4\,\simeq $ $
\,2\times 10^{-6}\,m_U(\hbox{MeV})\ {r}/{\sin 2\beta}\,$. 
We discuss phenomenological implications for meson 
decays, neutrino interactions,  atomic-physics parity violation, naturally suppressed $\pi^0\!\to\gamma \,U$ decays, etc.. The $U$ boson fits within the grand-unification framework, in symbiosis with a $SU(4)_{\rm es }$
{\em electrostrong\hspace{.2mm}} symmetry broken at the GUT scale, with $Q_U$  depending on
$Q,\,B\!-\!L,\ F_A$ and $T_{3A}$ through three parameters $\gamma_Y,\,\gamma_A$ and $\eta$\,.
\\
\end{abstract}

\maketitle

\section{\boldmath The general\vspace{1.5mm} picture \hbox{\ for a light spin-1 $\,U$ boson}}

\vspace{-1mm}

\subsection{A new light gauge boson}

\vspace{-2.5mm}

Could there be a neutral light gauge boson in nature, other than the photon? 
We have long discussed the possible existence of such a new boson called the $U$ boson, 
with weak or very weak axial and/or vector couplings to quarks and leptons \cite {U}. The vector part in the $U$ current is generally obtained, after mixing with the $Z$ current, as a combination of the conserved $B,\ L_i$ and electromagnetic currents~\cite{U2}. An axial part may be present  when the electroweak breaking is generated by two Brout-Englert-Higgs doublets at least, with different gauge quantum numbers. 
The new gauge boson then behaves very much, in the low mass limit, as an axionlike particle, which may be turned into mostly an electroweak singlet so as to be very weakly interacting \cite{U}. One can search for it in $e^+e^-$ annihilations,
$K$, $\psi$ and $\Upsilon$ decays,  beam dump experiments, neutrino scatterings, parity-violation effects, etc.. It may also serve as a mediator for the annihilation of dark matter particles, providing sufficient annihilations to allow for dark matter to be light \cite{ldm1,ldm2}, circumventing the Hut-Lee-Weinberg mass bound of a few GeV's \cite{hut,lw}.
It may contribute to ano\-ma\-lous magnetic moments of charged leptons, providing potential interpretations for possible deviations from their standard model (SM) expected values \cite{prd07}.
The search for such light bosons, very weakly coupled to  standard model particles and providing a possible bridge  to a new dark sector, has developed into a subject of intense experimental interest \cite{rev1,rev2}.

\vspace{1.5mm}

We would like to discuss here the general situation of a \hbox{spin-1} $U$ boson and how, by coupling to a combination of  the extra-$U(1)$ and $Z$ currents, it interpolates between a dark photon or generalized dark photon coupled to $Q,\,B$ and $L_i$ currents, a dark $Z$ coupled to the $Z$ current, and a light boson with axial couplings. The quark and lepton couplings are expressed in general as a linear combination of $Q,\,B,\,L_i,\,T_{3L}$ and $F_A$, satisfying, for one generation, two independent relations.
Axial couplings, with in general isovector and isoscalar contributions, are also at the origin of enhanced interactions for a longitudinal $U$ boson.

\vspace{-1mm}

\subsection{\boldmath An axionlike behavior of a light $U$ boson}

\vspace{-2mm}
The longitudinal polarisation state of a light $U$ boson with axial couplings undergoes enhanced interactions, by a factor $\approx k^\mu/m_U$ which may be large for small $m_U$. In fact the $U$ does not decouple even if its gauge coupling $g"$ gets very small, its mass $m_U$ getting very small as well,
its longitudinal polarisation state interacting with amplitudes 
$\propto g"\,k^\mu/m_U$ having finite limits. It behaves then very much as the corresponding Goldstone boson, a \hbox{spin-0} axionlike particle $a$, but it has no $\gamma\gamma $ decay mode. Its effective pseudoscalar couplings to quarks and leptons $g_P$ are obtained from its axial ones $g_A$ through the correspondence

\vspace{-6mm}
\be
g_A\ U_\mu \,\bar f\,\gamma^\mu\gamma_5\, f\ \ \to\ g_A\ \frac{2m_f}{m_U}\ a\ \bar f\,\gamma_5\,f\,,
\ee
so that \cite{U}

\vspace{-8mm}
\be
\label{gpga}
g_{P\,ql}\,= \,g_{A\,ql}\ \frac{2m_{ql}}{m_U}\ .
\ee 

The resulting enhancement of $U$ interactions for small $m_U$ could lead to too large effects.
One way to avoid them
 is to diminish these effective pseudoscalar couplings by introducing, next to the two doublets 
$h_1$ and $h_2$ considered in this case,  a \hbox{spin-0} singlet $\sigma$, now often known as a ``dark \hbox{Higgs''} field; it acquires a sufficiently large v.e.v. as compared to the electroweak scale, so that the extra $U(1)$ symmetry gets broken ``at a large scale''  \cite{U}. This turns the equivalent axionlike pseudo\-scalar from an active field $A$
into a mostly inert one $a$, very weakly interacting or nearly ``invisible''. It is expressed in terms of its doublet and singlet components as  

\vspace{-5mm}
\be
\label{rtheta}
a\,=\,\cos\theta_A \,A + \sin\theta_A \ (\sqrt 2 \ \hbox{Im}\ \sigma)\,,
\ee
its couplings to quarks and leptons being proportional to the invisibility parameter \vspace{0mm}
\be
r= \cos\theta_A <1\,. 
\ee
Its production rates
are proportional to $\,r^2\!=\cos^2 \theta_A\,$, with in particular, when the axial generator $F_A$ participates in the gauging so that a longitudinal $U$ boson gets produced much like an invisible axion,
\be
\label{psi}
\ba{l}
\hspace{-3mm}B\,[\,\psi \,(\Upsilon) \to \gamma + U\,]\,\simeq
\vspace{2mm}\\
\hspace{5mm} r^2\ B\,[\,\psi \,(\Upsilon)\to \gamma 
+ \hbox{standard axion}\,]\,, \ \ \hbox{etc.}.\hspace{-5mm}
\ea
\ee
\noindent 
These rates are sufficiently small if $r=\cos\theta_A$ is small enough, i.e. if the pseudoscalar $a$, turned into the longitudinal degree of freedom of the light $U$ boson, is mostly an electroweak singlet.

\vspace{-1.5mm}

\subsection{\boldmath The $U$ current, combination\vspace{1mm} of the extra-$U(1)$ \hbox{and $Z$ currents}}

\vspace{-1mm}

In this approach, initiated forty years ago, the $SU(3) \times SU(2)\times U(1)_Y$ gauge group of the standard model (SM) is extended to include an extra $U(1)_F$ factor. Its gauge field $C^\mu$, taken to be very weakly coupled, mixes in general with the $Z_\circ^\mu$ field of the standard model, according to
\vspace{-3mm}
\be
\label{zumix}
U^\mu =\, \cos\xi \ C^\mu + \sin\xi \ Z_\circ^\mu  \,,
\ee
with a small mixing angle $\xi$ given by \cite{U2}
\be
\label{xi0}
\tan\xi \,\simeq\,\frac{g"}{\sqrt{g^2+g'^2}}\ \ {\sum_i\, \epsilon_i\, F(h_i)\ \frac{v_i^2}{v^2}}\,
=\,\frac{g"}{g_Z}\ \Gamma\,.
\ee

\vspace{1mm}

The mixing  is determined by the extra $U(1)_F$ coupling $g"$, and by a sum on the Brout-Englert-Higgs doublets $h_i$, denoted by $\Gamma\,$.
The $h_i$'s are doublets with weak hypercharges $Y(h_i)=\epsilon_i=\pm 1$ and v.e.v.'s $\,v_i/\sqrt 2$, 
\vspace{-.2mm}
contributing to the mixing with the $Z$, 
\vspace{-.3mm}
with  $v^2=\sum_i v_i^2$ $=  1/(G_F\sqrt 2)  \simeq (246.22\ \hbox{GeV})^2$.
Extra singlets, when present, can contribute significantly to the $U$ mass but not
to its mixing with the $Z$, in the small mass limit.
The $U$ current is thus obtained in a general way as the following linear combination of the extra-$U(1)_F$ and $Z$ currents,
\bea
\label{juold}
\hspace{-4mm}
\framebox [8.6cm]{\rule[-.8cm]{0cm}{2cm} $ \dis
\ba{ccl}
{\cal J}^\mu_U&\simeq &
 \, g"  \cos\xi \ \dis \times  
\vspace{2mm}\\
&&\hspace{-8mm} \hbox{\LARGE$\left[\right.$}\ \hbox{$\dis \frac{1}{2}$}\ J^\mu_F \, +
\, \underbrace{\hbox{\large $\left(\right.$} \hbox{$ \sum_i$} \  \epsilon_i \,F(h_i) \ \dis {v_i^2}/{v^2} \hbox{\large $\left.\right)$}}_{\hbox{\small $\Gamma$}}\,  
 (J_{3L}^\mu-\sin^2 \theta \,J_{\rm em}^\mu)\,  \hbox{\LARGE$\left.\right]$}.
\ea
$}\hspace{-8mm}
\nonumber \vspace{-5mm}\\
\eea
In the situations usually considered here the mixing angle $\xi$ is small so that $\cos\xi\simeq 1\,$.

\vspace{2mm}
Two classes of models are particularly interesting and representative.
With $F$ taken as the weak hypercharge $Y$ (plus a possible dark matter contribution),  a single doublet  as in the standard model (or two as in the supersymmetric standard model, with $F\!=Y\!=1$ for $h_2$ and $h_1^c$) so that $\Gamma=1$, and an extra singlet $\sigma$ to generate the $U$ mass, the $\frac{1}{2}\,J^\mu_Y$ and $J^\mu_{3L}$ terms in eq.\,(\ref{juold}), both parity-violating, recombine into $J_{\rm em}^\mu$.
{\it The mixing of the $Y$ current with the $Z$ current reconstructs exactly the electromagnetic current,}
with
\be
\label{juold1}
{\cal J}^\mu_U\,\simeq\,
 \,\dis g"\cos\xi \,  \cos^2\theta \ J_{\rm em}^\mu\,+\,{\cal J}^\mu_{U\,d}\,.
\ee
The $U$ boson appears as a ``dark photon'' with a coupling $g" \cos\xi   \cos^2\theta = e\,\tan\chi$ to the electromagnetic current. When $F$ includes also, in addition to $Y$, 
terms proportional to baryon and lepton numbers, the $U$ appears as a generalized dark photon coupled to a linear combination of $Q,\,B$ and $L_i$ currents \cite{U2,epjc},
\be
\label{juold1bis}
{\cal J}^\mu_U\simeq
 \dis g" \cos\xi  \left[ \cos^2\theta \, J_{\rm em}^\mu+\frac{1}{2}\,(\gamma_BJ^\mu_B+\gamma_{L_i}J^\mu_{L_i})\right] 
 \!+ {\cal J}^\mu_{U\,d}\,.
\ee

\vspace{2mm}

In contrast $F$ may be taken as an axial symmetry generator $F_A$
(equal to $- 1/2$ and $+1/2$ for left-handed  and right-handed quark and lepton fields, respectively), 
plus a possible dark matter contribution. This requires two BEH doublets at least, taken as $h_1$ and $h_2$ with $\tan\beta=v_2/v_1$ as in supersymmetric extensions of the standard model   \cite{plb76}. These two doublets, taken with $Y=\mp 1$ and $F_A=1$, may then be rotated independently thanks to $U(1)_A$ and $\,U(1)_Y$. We thus have, as seen from (\ref{juold}) with $\Gamma = (v_2^2-v_1^2)/v^2=-\cos 2\beta$ \cite{plb86}, 
\be
\label{juold2}
{\cal J}^\mu_U\simeq
 \,\dis g" \cos\xi \  \hbox{\LARGE$\left[\right.$}\,\hbox{$\dis \frac{1}{2}$}\ J^\mu_A \,
-\,\cos 2\beta \,(J_{3L}^\mu-\sin^2 \theta \,J_{\rm em}^\mu) \hbox{\LARGE$\left.\right]$}+{\cal J}^\mu_{U\,d}\,.
\ee
Or, including $B$ and $L_i$ contributions as in (\ref{juold1bis}), 
\be
\label{juold2bis}
\ba{ccc}
{\cal J}^\mu_U\,\simeq\,
 \,\dis g" \cos\xi \  \hbox{\LARGE$\left[\right.$}\,\hbox{$\dis \frac{1}{2}$}\ J^\mu_A \,
-\,\cos 2\beta \,(J_{3L}^\mu-\sin^2 \theta \,J_{\rm em}^\mu) \hbox{\LARGE$\left.\right]$}
\vspace{2mm}\\
\dis\, +\ g"\cos\xi \ \ \frac{1}{2}\ (\gamma_BJ^\mu_B+\gamma_{L_i}J^\mu_{L_i})\,+\,{\cal J}^\mu_{U\,d}\ .\  
\vspace{-5.2mm}\\
\ea
\ee

\vspace{5mm}

Doublet BEH fields uncoupled to quarks and leptons but with non-vanishing v.e.v.'s, referred to as ``semi-inert'', may also participate in the mixing, again with $C^\mu$ very weakly coupled to quarks and leptons through a linear combination of the $Y,\,B,\,L_i$ and $F_A$ currents.
As a special case it is even possible that the $U(1)_F$ gauge field $C^\mu$ does not couple to quarks and leptons at all,
while still mixing with $Z^\mu_\circ$ as in (\ref{zumix},\ref{xi0}) thanks to the v.e.v.'s of semi-inert doublets transforming under $U(1)_F$. The resulting current for quarks, leptons and dark matter 
is then obtained from (\ref{zumix}-\ref{juold}) as
\be
\label{juold3}
\ba{ccl}
{\cal J}^\mu_U \!&=&\! \sin\xi\ \sqrt{g^2+g'^2}\ (J_{3L}^\mu-\sin^2 \theta \,J_{\rm em}^\mu) \,+\,{\cal J}^\mu_{U\,d}
\vspace{2mm}\\
&&\hspace{-11mm}\simeq\, g"\cos\xi \
\hbox{\Large $\left(\right.$} \hbox{$\sum_i$} \, \epsilon_i \,F(h_i) \ \dis \frac{v_i^2}{v^2} \hbox{\Large $\left.\right)$} 
\, (J_{3L}^\mu-\sin^2 \theta \,J_{\rm em}^\mu) 
+\,{\cal J}^\mu_{U\,d}\,.
\ea
\ee
Only semi-inert doublets contribute to $\Gamma$ when the other doublets coupled to quarks and leptons have $F=0$,
 a semi-inert doublet $h'$ taken with $F\!=2$ 
leading to $\,\Gamma= $ $2 \,v'^2/v^2= 2 \sin^2\beta'$ in (\ref{juold3}).
Such a $U$ boson simply coupled to the $Z$ weak neutral current  may be referred to as a \hbox{dark $Z$} boson \cite{mar,marbis,jung}.
Again, other contributions proportional to $Q,\,B,\,L_i$ and $F_A$ currents may be present as in (\ref{juold1bis},\ref{juold2bis}).

\vspace{2mm}
For $\Gamma=0$, $\ C^\mu$ and $Z^\mu_0$ do not mix, the $U$ 
current for quarks, leptons and dark matter being simply given by
\vspace{-1mm}
\be
\label{unmix}
{\cal J}^\mu_U \,=\,\frac{g"}{2}\, \left(\,\gamma_A J^\mu_A + \gamma_Y\, J^\mu_Y + \gamma_B \,J^\mu_B+\gamma_{L_i} J^\mu_{L_i}\,\right)+\,{\cal J}^\mu_{U\,d}\,.
\ee
\vspace{-6mm}

\subsection{\boldmath The $U$ charge \vspace{1.5mm} as a general combination of $Q,\,B,\,L_i,\ T_{3L}$ and $F_A$}

\vspace{-1mm}

We shall discuss more precisely in Section \ref{sec:u1} how the
extra-$U(1)_F$ symmetry to be gauged may be generated by an arbitrary  linear combination of the axial generator $F_A$ with $Y,\,B$ and $L_i$, plus a dark matter contribution, and a possible $\gamma_{F'}F'$ term acting on semi-inert BEH doublets uncoupled to quarks and leptons. This extra-$U(1)$ generator is then given by 
\be
\label{F-1}
\!F\,=\, \gamma_A F_A + \gamma_Y Y + \gamma_B B+\gamma_{L_i} L_i + \gamma_d \,F_d+\gamma_{F'}F'\,.
\ee
This includes specific combinations for which $U(1)_F$ does not act on left-handed quarks and leptons but only on right-handed ones, such as 
\be
\label{t3r}
T_{3R}=\frac{Y}{2} -\,\frac{B-L}{2}\,,
\ee
or $ (3B+L)/2+ F_A$,
equal to $+\,1$ for  right-handed quark and lepton fields.

\vspace{2.5mm}

For two active doublets $h_1$ and $h_2$  with $Y=\mp \,1$ and $F_A=1$, plus a singlet $\sigma$, the mixing angle $\xi$ in (\ref{xi0}) is given by
\vspace{-1.5mm}
\be
\label{ximod}
\tan \xi\,\simeq\,\frac{g"}{\sqrt{g^2+g'^2}}\ (\gamma_Y-\gamma_A\,\cos 2\beta)\,.
\ee
Remarkably, it allows us 
to unify within a single formula the situations described in (\ref{juold1}-\ref{juold2bis}), 
with vector and/or axial couplings of the $U$ boson, by expressing  the $U$ current in (\ref{juold}) as the sum of three contributions:
\vspace{1mm}

1) a vector current linear combination of $Q,\,B$ and $L_i$  currents, as for a generalized dark photon in (\ref{juold1bis})
\cite{U2,epjc};

\vspace{1mm}

2) a combination of a universal axial current $J^\mu_A$ 
\vspace{-.3mm}
with the standard weak neutral current
$J^\mu_{3L} -\,\sin^2\!\theta\,J^\mu_{\rm em}$ as in (\ref{juold2}), as for an axial boson mixed with the $Z$ 
 \cite{plb86};
 
 \vspace{1mm}
 
3) and a dark matter contribution \cite{ldm1,ldm2}. 

 \vspace{1.2mm}
 
 \noindent
This reads:

\vspace{-2.5mm}

\bea
\label{juold5}
\ba{ccl}
{\cal J}_U^\mu\,\simeq \ \ \ 
g"\cos\xi\  \left[\,\gamma_Y\cos^2\theta\,J_{\rm em}^\mu +\frac{1}{2}\,(\gamma_B J^\mu_B + \gamma_i\,J^\mu_{L_i}) \, \right]
\vspace{3.5mm}\\
\hspace{9mm} +\ g"\cos\xi \ \gamma_A \left[\,\frac{1}{2} \,J^\mu_{A}\!-\,\cos 2\beta\, (J^\mu_{3L}-\sin^2\theta J^\mu_{\rm em}  )\,\right]
 \vspace{2.5mm}\\
\dis  +\ \frac{g"}{2}\,\cos\xi \ \gamma_d\,J^\mu_d\,.
\ea 
\nonumber
\eea
\vspace{-14.5mm}
\be
\ee

\vspace{2mm}

In a general way we can decompose $F$ as  $\gamma_Y Y\!+\! \tilde F$, so that
\vspace{-2mm}
\be
\label{delta0}
\Gamma\,=\, \gamma_Y + \hbox{\large$\sum_i$}\ \epsilon_i \tilde F(h_i)\ \frac{v_i^2}{v^2}\,=\,\gamma_Y+\,\eta\,,
\ee
with
\vspace{-6mm}

\be
\label{eta}
\eta\,=\, \hbox{\large$\sum_i$}\ \epsilon_i\, [\,\gamma_A F_A+\gamma_{F'} F'\,](h_i)\ {v_i^2}/{v^2}\, ,
\ee
\vspace{-3mm}

\noindent
leading to

\vspace{-7mm}
\be
\label{txieta}
\tan\xi\,=\,\frac{g"}{g_Z}\ (\gamma_Y+\eta)\,,
\ee
as in (\ref{ximod}).
With the usual active doublets $h_1,h_2$ ($Y=\mp\,1,\ F_A\!=1$), a semi-inert one $h'$ (taken for future convenience with $F'=2\,Y=2$), 
and a singlet $\sigma$ not contributing to $\eta$, we have
\vspace{-4mm}

\be
\label{etabis}
\ba{ccl}
\eta\!&= &\ \dis\gamma_A\,\frac{v_2^2-v_1^2}{v^2}+2\,\gamma_{F'}\,\frac{v'^2}{v^2}
\vspace{2mm}\\
&=&\!\dis -\,\gamma_A\,\cos2\beta\, \cos^2\!\beta'+2\,\gamma_{F'}\,\sin^2\beta'\,.
\ea
\ee
This reduces to  $\eta= -\,\cos 2\beta\,$ 
for $\,h_1,h_2,\,\sigma$ (choosing $\gamma_A=1$), and
$\eta=2\,{v'^2}/{v^2}=2\sin^2\!\beta'$
for a SM-like doublet $h_{\rm sm}$ and a semi-inert one $h'$ (choosing $\gamma_{F'}=1$), with a possible singlet.

\vspace{2mm}

 The $U$ current is then expressed as
 \be
\label{curr40}
\ba{l}
\!\!{\cal J}_U^\mu\,\simeq\,g"\,\cos\xi\ \times
\vspace{2mm}\\
\hspace{3mm}\dis\hbox{\Large$\left[\right.$}\,  \frac{1}{2}\, (\gamma_A J^\mu_{A}+ 
\gamma_Y J^\mu_Y +  \gamma_B J^\mu_B + \gamma_i\,J^\mu_{L_i} \!+\gamma_{F'}J^\mu_{F'}+ \gamma_d J^\mu_d)
\vspace{0mm}\\
 \dis\hspace{19mm} +\ (\gamma_Y +\eta) \, (J_{3L}^\mu-\sin^2 \theta \,J_{\rm em}^\mu)\, 
 \hspace{-3mm}\phantom{\frac{1}{2}} \hbox{\Large$\left.\right]$}\ .
 \vspace{-5.5mm}\\
\ea
\ee
\vspace{0mm}

\noindent
Recombining $\gamma_Y\,(\frac{1}{2} \,J^\mu_Y + J_{3L}^\mu-\sin^2\theta\, J^\mu_{\rm em})$ into $\,\gamma_Y \cos^2\theta$ $J^\mu_{\rm em}$, we find the general expression of ${\cal J}^\mu_U$ as
\be
\label{juold4-1}
\framebox [8.6cm]{\rule[-1.5cm]{0cm}{2.6cm} $ \dis
\ba{cl}
{\cal J}_U^\mu&\simeq 
\dis \ g"\cos\xi\  \dis\hbox{\Large$\left[\right.$}\ \gamma_Y\cos^2\theta\,J_{\rm em}^\mu +\frac{1}{2}\,(\gamma_B J^\mu_B + \gamma_i\,J^\mu_{L_i}) \ \dis\hbox{\Large$\left.\right]$}\,
\vspace{1.5mm}\\
&\dis\  +\ g"\cos\xi \ \dis\hbox{\Large$\left[\right.$}\ \frac{\gamma_A}{2} \,J^\mu_{A}+\,\eta\, (J^\mu_{3L}-\sin^2\theta J^\mu_{\rm em}  )\ 
\dis\hbox{\Large$\left.\right]$}
 \vspace{1mm}\\
&\dis \ +\ g"\cos\xi \ \ \  \frac{1}{2}\ (\gamma_{F'} J^\mu_{F'} + \gamma_d\,J^\mu_d)\ .
\vspace{-5.5mm}\\
\ea
$}
\ee

\vspace{1mm}

\noindent
It involves a general linear combination of {\it electromagnetic and $Z$ currents with baryonic and leptonic currents, and  an axial current,} with an additional dark matter contribution. 

\vspace{2mm}
The mixing with the $Z$ current is determined by the parameters 
$\gamma_Y$ and $\eta$ in (\ref{delta0}-\ref{etabis}).
For $\gamma_A=\eta=0$ we recover a $U$  vectorially coupled to $Q,\,B$ and $L_i$.
For $\gamma_A\!\neq 0$ 
we recover the $U$ current
 as an axial one mixed with the $Z$ current and combined with $Q,\,B$ and $L_i$ currents.
For $\gamma_A\!=0\,$, $\gamma_Y$ and $\eta\neq 0\,$ we get a mixing between a dark photon and a dark $Z$, also coupled to baryonic and leptonic currents. 
\vspace{2mm}

This corresponds to  a $U$ charge, associated with the coupling $g"\cos\xi$ (in most cases close to $g"$), obtained from the $Z$-$U$ mixing in (\ref{zumix}) as
\be
\label{qufz}
Q_U\,=\,\dis \frac{1}{2}\ F\,+\,\tan\xi\ \,\frac{g_Z}{g"}\ (T_{3L}-\sin^2\theta\,Q)\,.
\ee
It reads in the small $m_U$ limit
\be
\label{qufz2}
\ba{ccc}
Q_U\,\simeq\,\dis \frac{1}{2}\,\left[\,\gamma_AF_A+\gamma_YY\!+\gamma_B B+\gamma_{L_i}L_i\!+\gamma_{F'} F'\!+\gamma_d F_d
\,\right]
\vspace{2mm}\\
+\ (\gamma_Y+\eta)\ \,(T_{3L}-\sin^2\theta\ Q)\,,
\vspace{-4mm}\\
\ea
\ee

\vspace{4mm}
\noindent
reexpressed as
\be
\label{qufz2}
\!\!\framebox [7.8cm]{\rule[-.85cm]{0cm}{1.9cm} $ \dis
\ba{ccc}
\!\!\dis Q_U\simeq\,\gamma_Y\cos^2\theta \ Q + \frac{1}{2}\,(\gamma_AF_A+\gamma_B B+\gamma_{L_i}L_i )
\vspace{2mm}\\
+ \ \eta\ (T_{3L}-\sin^2\theta\ Q)\,+\,\dis \frac{1}{2}\,(\gamma_{F'} F'\!+\gamma_d\,F_d)\,,
\ea
$}\!\!\!
\ee
in agreement with expression (\ref{juold4-1}) of $\,{\cal J}^\mu_U\,$.

\vspace{-.5mm}

 \subsection{\boldmath Special cases of a $U$ unmixed with the $Z$}
 \vspace{-1mm}
 
When $\Gamma\!=\gamma_Y\!+\eta=0\,$ so that $\xi\!=0$, there is no mixing between the standard-model $Z_0^\mu$ and the extra-$U(1) \ C^\mu$ gauge fields. $Z^\mu\!\equiv Z_0^\mu$ remains coupled to quarks and leptons as in the SM, and $U^\mu\!\equiv C^\mu$ is coupled to the extra-$U(1)$
current $J^\mu_F\,$ as in (\ref{unmix}). This is the case, in particular, for an axial gauge boson unmixed with the $Z$ 
(with $\gamma_Y\!=0,\,\gamma_A\!=1$, and $ \eta= -\cos 2\beta=0$  for $v_1=v_2$), which may also be coupled to $B$ and $L_i$ \cite{U}. 

\vspace{2mm}
 This may also occur with a semi-inert doublet $h'$, for $\gamma_A=0$ and $ \Gamma=\gamma_Y+2\gamma_{F'}\sin^2\!\beta'\!=0$\,. As a very special case we may take $\gamma_Y\!=-\,\gamma_{F'}=1$ with SM-like and semi-inert doublets $h_{\rm sm}$ and $h'$ having opposite $F=\pm 1$ and equal v.e.v.'s  ($\tan\beta'=1$) so that $C^\mu$ and $Z^\mu_0$ do not mix. In all such situations with no mixing we get from (\ref{curr40}) the $U$ current
 \be
 \label{nomix}
\ba{l}
\!\!\!{\cal J}_U^\mu\,=\,\dis \frac{g"}{2}\,J^\mu_F \,=\ \dis \frac{g"}{2}\ \times
\vspace{3mm}\\
\hspace{7mm}\dis (\gamma_A J^\mu_{A}+ 
\gamma_Y J^\mu_Y +  \gamma_B J^\mu_B + \gamma_i\,J^\mu_{L_i} \!+\gamma_{F'}J^\mu_{F'}+ \gamma_d J^\mu_d)\ .
\vspace{1mm}\\
\ea
\ee
It corresponds to
\be
\label{nomix2}
Q_U=\frac{F}{2}=\frac{1}{2}\ (\gamma_A F_{A}+ 
\gamma_Y Y +  \gamma_B B + \gamma_i\,{L_i} \!+\gamma_{F'} {F'}+ \gamma_d F_d)\,,
\ee
acting on quarks and leptons as a linear combination of $Y,\,B$ and $L_i$ (including possibly $T_{3R}$) with the axial $F_A$\,. This unmixed expression of $Q_U=F/2\,$  is also easily recovered from the general expression  (\ref{qufz2}), under the no-mixing  constraint  $\,\eta =- \,\gamma_Y$\,.

\vspace{-2mm}

\subsection{\boldmath Vector and axial parts in the $U$ charge}

\vspace{-1mm}
Returning to the general situation of the $U$ current in (\ref{curr40}) the couplings of chiral quark and lepton fields are expressed as 
$
g"\cos\xi\ Q_U$,
with $Q_U$ given in (\ref{qufz2}) as a  linear combination of
$Q,\,B,\,L_i,\, T_{3L}$ and $F_A$
(the colorless neutral symmetry generators in the standard model and its extensions)\,\, \footnote{From now on we shall often use $\,=\,$ instead of $\,\simeq\,$ for simplicity, for approximate equalities becoming exact in the small $m_U$ limit.}:
\be
\label{coup0}
\ba{c}
\dis Q_U\,=\,(\gamma_Y\cos^2\theta -\eta\,\sin^2\theta)\ Q\,  +\,\frac{\gamma_B}{2}\ B\,+\frac{\gamma_{L_i}}{2}\ L_i\ 
\vspace{1.5mm}\\
\hspace{20mm}\dis \,+\,\frac{\gamma_A}{2}\ F_A\,+\,\eta\ T_{3L}\,.
\vspace{-5.5mm}\\
\ea
\ee
\vspace{-1mm}

\noindent
The couplings are expressed in terms of {\it \,five independent parameters\,} if we assume lepton universality, $\gamma_Y$, 
$\gamma_B$  and $\gamma_L$, $\gamma_A$ and $\eta$, all of them multiplied by $g"\cos\xi\,$.
This allows for many interesting situations, with in particular the possibility of reduced couplings to neutrinos, and to protons.
In addition to vector couplings involving $Q,\,B$ and $L_i$, we also get in general axial couplings, involving isoscalar as well as isovector contributions.

\vspace{2mm}
Let us
 separate the parts corresponding to vector and axial couplings, with $\,Q=T_{3L}+T_{3R}+(B-L)/2$ so that
\be
\ba{ccc}
T_{3L}-\sin^2\!\theta\ Q\!&=&\! \dis\frac{T_{3L}+T_{3R}}{2}\,-\,\sin^2\theta\,Q -\,\frac{T_{3R}-T_{3L}}{2}
\vspace{2mm}\\
\!&=&\! \dis (\,\frac{1}{2}-\sin^2\theta\,) \ Q -\frac{1}{4}\,(B-L)-\,\frac{T_{3A}}{2}\,,
\ea
\ee
\noindent
and
\be
\label{quv}
\hspace{-1mm}\framebox [7.85cm]{\rule[-.85cm]{0cm}{2.5cm} $ \dis
\ba{ccl}
\hspace{-2mm}(Q_U)_V\!\!&=&\gamma_Y\cos^2\theta \ Q\,  +\,\dis\frac{\gamma_B}{2}\ B\,+\frac{\gamma_{L_i}}{2}\ L_i
\vspace{1.5mm}\\
&& \!\dis +\ \eta\ \hbox{\Large$\left[\right.$} (\,\frac{1}{2}-\,\sin^2\theta\,)\ Q  -\,\frac{1}{4}\, (B-L)\hbox{\Large$\left.\right]$}\, ,\!\!
\vspace{2.5mm}\\
\hspace{-2mm}(Q_U)_A\!\!&=&\dis \frac{\gamma_A}{2}\,F_A  -\,\frac{\eta}{2}\ T_{3A}\,.
\vspace{2mm}\\
\ea
$}\!\!
\ee
\vspace{1mm}

With $T_3=T_{3L}+T_{3R}$, $T_{3A}=-T_{3L}+T_{3R}$ and $T_{3L}= (T_3-T_{3A})/2$,
one can also write, if one prefers,  $\,(T_{3L})_A = -\,T_{3A}/2$\,, and $(T_{3L})_V=T_3/2=Q/2-(B-L)/4\,$.
Let us consider as a check $\,\eta=-\,\gamma_Y$ so that there is no mixing with the $Z$. Then
\be
\ba{c}
\left\{ \
\ba{ccl}
(Q_U)_V\!&=&\!\gamma_Y\, [\,Q-(T_{3L})_V\,]+\,\dis\frac{\gamma_B}{2}\ B\,+\frac{\gamma_{L_i}}{2}\ L_i\,,
\vspace{1mm}\\
(Q_U)_A\!&=&\!\ \ \dis\frac{\gamma_A}{2}\,F_A-\gamma_Y\,(T_{3L})_A\,.
\ea\right.
\vspace{-5mm}\\
\ea
\ee

\vspace{3.5mm}

\noindent
Summing the two we get for the quark and lepton couplings, with $Q-T_{3L}=Y/2$, 
 \be
Q_U = \frac{\gamma_A}{2}\,F_A+\frac{\gamma_Y}{2}\,Y\!+\dis\frac{\gamma_B}{2}\, B+\frac{\gamma_{L_i}}{2}\, L_i
\,= \,\frac{1}{2}\, F\,,
 \ee
 which provides back $Q_U=F/2$ as in (\ref{qufz2},\ref{nomix2}), in the absence of mixing with the $Z$.

\vspace{-1.5mm}

\subsection{More on the vector couplings}

\vspace{-1.5mm}

The vector couplings involve a linear combination of the conserved $B,\ L_i$ and electromagnetic currents.
The pure dark photon case corresponds to the {\it specific direction $(1;\,0, \,0,\!0,\!0\,;\,0,\,0)$ in the seven-dimensional 
parameter 
\vspace*{-3mm}
space}
\be
(\,\gamma_Y;\,\gamma_B,\,\gamma_{L_i};\,\gamma_A,\,\eta\,)\,.
\ee

\vspace{1mm}

As a result of the mixing with $Z$ we recognize in the vector contribution to the $U$ charge the part corresponding to the weak charge \cite{plb86,bf}, defined as  
\be
\label{qzu}
(Q_Z)_V\!=  (T_{3L})_V \,-\,\sin^2 \theta \ Q =(\,\frac{1}{2}\,-\,\sin^2\theta)\, Q\,-\,\frac{1}{4}\ (B-L)\, .
\ee
When conventionally  renormalized by a factor $4$ into  
\be
\label{qzu2}
Q_{\rm \,Weak}= 4\,(Q_Z)_V= (2-4\,\sin^2\theta))\ Q-(B-L)\, ,
\ee
it reduces to the usual definition of the weak charge for a nucleus, $Q_{\rm \,Weak}(Z,N)= (1-4\sin^2\theta)\,Z\,-N$  \cite{boubou}.

\vspace{2mm}

For $\sin^2\theta$ having the grand-unification value 3/8 \cite{gg}
the vector couplings of the $U$ are fixed by
\be
\label{qzgut}
Q_{Z\,V}^{\rm\, gut}\,=\,\frac{1}{8}\ Q-\frac{1}{4} \ (B\!-\!L)\,.
\ee
They vanish for up quarks, as do the $Z$ vector couplings to up quarks. We shall come back to this in Section \ref{sec:gut},
in connection with a $SU(4)_{\rm es}$ {\it electrostrong} symmetry relating the photon with the eight gluons.

\vspace{2mm}
With $\sin^2\theta$ close to 1/4\, however, the $Q_Z$ contribution to the vector couplings in (\ref{qzu}), close to be proportional to
$B-L-\,Q\,$ (rather than to $\,B-L-\,Q/2\,$ for $\sin^2\!\theta=3/8$),
 is significantly smaller for the proton than for the neutron  \cite{plb86}, and thus naturally ``protophobic'', instead of being ``$u$-phobic'' at the GUT scale.
 
 \vspace{-1.5mm}
 
\subsection{\boldmath $P,\ C$ and $CP$ properties of the $U$ boson}

\vspace{-1.5mm}

There are different representatives within the general class of models considered, associated with different choices  for the gauged extra-$U(1)$ quantum number $F$ in (\ref{F-1}),
the doublet and singlet v.e.v.'s,  etc..
The resulting $U$ current in (\ref{juold}-\ref{unmix},\ref{juold5}) or (\ref{juold4-1}) may be purely {\em vectorial}, purely {\em axial}, or in general appears as {\em a combination of vector and axial parts}, with 
\be
\label{cp}
\!\!\left\{\ba{cl}
\hbox{vectorially-coupled}\ U: & \!CP^{--} \   \hbox{(as the photon)}, \!\!
\vspace{2mm}\\
\hbox{axially-coupled}\ U: & \!CP^{++} \,.
\ea\right.
\ee
If both couplings are present the $U$ interactions are $C$ and $P$-violating, the $U$ still having $CP= \,+\,$. 
Its interactions generally preserve $CP$ except for very small effects, including a possible very small $CP$-violating monopole-dipole spin-dependent interaction \cite{cqg}.

\vspace{2mm}

Let us now concentrate on the axial couplings.

\vspace{-1mm}

\section{\vspace{1.8mm}Axial couplings
 \hbox{and effective pseudoscalar} ones}
\label{sec:genax}

\vspace{-1.5mm}

\subsection{Universal expressions of axial couplings}

\vspace{-2mm}

The rearrangements leading to expressions (\ref{juold5},\ref{juold4-1}) of the current may also be understood by considering the axial part in  the quark and lepton contribution to $\,{\cal J}^\mu_U$, 
\be
\label{jax}
({\cal J}_U^\mu)_{\rm ax}\,=\ g"\cos\xi \ \left[\,\frac{\gamma_A}{2} \,J^\mu_{A}\!+\,\eta\  (J^\mu_{3L})_{\rm ax} \,\right]\,,
\ee
which is {\it independent of} $\,\gamma_Y$,\, and obviously of $\gamma_B$ and $\gamma_{L_i}$.
It leads to the axial couplings $g_A=(g_R-g_L)/2\,$,
\be
\label{ga0bis0}
\ba{c}
\hspace{-6mm}\dis g_{A\pm}\,=\ g"\cos\xi \ (Q_U)_{A\pm} \,=\,\frac{g"\cos\xi}{4} \, \ [\,\gamma_A\,  \mp\,\eta\ ]\, ,
\vspace{-1mm}\\
\hspace{50mm}  \uparrow\hspace{7mm}\uparrow
\vspace{-.5mm}\\
\hspace{50.5mm}  \hbox{\small \it isoscalar\ \ \ isovector}
\vspace{-13.5mm}
\ea
\ee

\vspace*{10.5mm}

\noindent
independently of $\gamma_Y,\gamma_B$ and $\gamma_{L_i}$. It corresponds to $(Q_U)_A$ in (\ref{quv}), with
the $\gamma_A$ and $\mp\ \eta$ terms providing {\it isoscalar} and {\it isovector\,} contributions, the latter induced by the mixing with the $Z$.
All axial couplings are given by the two universal expressions
\be
\label{gaeb20}
 \framebox [8.2cm]{\rule[-.9cm]{0cm}{2cm} $ \dis
\left\{\ba{rcccc}
g_{A+}\!\!&=&\!g_{A\,uct} \!&=&\dis \frac{g"\cos\xi}{4}\,\ (\gamma_A-\eta)\ ,
\vspace{2mm}\\
g_{A-}\!\!&=&\!g_{A\,e\mu\tau}= \,g_{A\,dsb}\!&=&\dis \frac{g"\cos\xi}{4}\,\ (\gamma_A+\eta)\ .
\ea\right.
$}
\ee

\vspace{3mm}

They satisfy the general universality relations
\be
\label{univ}
\left\{\ \ba{ccc}
g_{Au}=g_{Ac}= g_{At}\,,
\vspace{2mm}\\
g_{Ae}=g_{A\mu}= g_{A\tau}\,= \, g_{Ad}=g_{As}=g_{Ab}\,.
\ea \right.
\ee 
These are in fact a consequence of the hypothesis that up quarks on one hand, down quarks and charged leptons on the other hand, all acquire their masses from the same BEH doublet, as in the standard model, or from $h_2$ and $h_1$ as in supersymmetric theories.
However, should we use more doublets and allow for different axial generators acting separately on leptons of the three families, and on quarks, to participate in the $U(1)_F$ gauging (cf. subsection\,\ref{subsec:further}), then the universality relations (\ref{univ})
and subsequent constraints on the axial $g_A$'s  would have to be abandoned.

\vspace{2mm}

For a purely isoscalar axial coupling or a purely isovector one, all axial couplings have the same magnitude
\be
\label{univuniv}
|\,g_{A\,uct}\,|
=|\,g_{A\,dsb}\,|=|\,g_{A\,e\mu\tau}\,|\,.
\ee
But the two active doublets $h_1$ and $h_2$ ($Y\!=\mp1,\,F_A\!=1$), \linebreak 
with the singlet $\sigma$, lead to a combination of isoscalar and isovector couplings.
Norma\-lizing $g"\!$ so that $\gamma_A=1$, $\eta=-\,\cos2\beta$, we recover \cite{plb86}
\be
\label{jaxbis}
({\cal J}_U^\mu)_{\rm ax}\,=\, g"\cos\xi \ \left[\,\frac{1}{2} \,J^\mu_{A}\!-\,\cos 2\beta\, (J^\mu_{3L})_{\rm ax} \,\right]\,,
\ee
providing  
 \be
\label{ga00}
\!g_{A\pm}\,=\,\frac{g"\cos\xi}{4}\ (1\,\pm\, \cos 2\beta)= \,\frac{g"\cos  \xi}{2} \,\times
\left\{\ba{c} \cos^2\beta\,, \vspace{2mm}\\ \sin^2 \beta\, . \ea\right. \!
\ee
The upper sign +  is relative to up quarks, and the lower one $-$ to down quarks and charged leptons.
\vspace{2mm}

All these expressions (\ref{jax}-\ref{ga00}) for the axial couplings are also compatible with grand-unification and the $SU(4)_{\rm es}$ electrostrong symmetry, as we shall see in Section \ref{sec:gut}.

\vspace{-1mm}

\subsection{\boldmath \hspace{-1.5mm}\boldmath\ $g"$ as proportional\vspace{1.5mm} to $m_U$ and to \hbox{\ \ \ \ \,the invisibility parameter $r$}} 

\vspace{-1mm}
Let us consider two BEH doublets $h_1^c$ and $h_2$ (or $h_{\rm sm}$ and $h'$) with different gauge quantum numbers, breaking spontaneously the $U(1)_U$ symmetry and making the $U$ massive, its current having an axial part. Its mass, initially given in the small $m_U$ limit by
\be
m_U\simeq\,\frac{g"\cos\xi\ v}{2}\ \sin 2\beta\,,
\ee
gets increased by the effect of an extra singlet v.e.v. $<\!\sigma\!>$ $=w/\sqrt 2$. It can then be expressed as \cite{plb86}
\vspace{1mm}
\be
\label{mumu}
m_U= \frac{g"\cos\xi}{2}\, \sqrt{v^2\sin^2 2\beta+F_\sigma^2w^2}\,=\,\frac{g"\cos\xi}{2} \,\frac{v\sin 2\beta}{r}\,,
\ee
with the invisibility parameter
\be
\label{r20}
r=\cos\theta_A\,=\,\frac{v\,\sin 2\beta}{\sqrt{v^2\,\sin^2 2\beta + F_\sigma^2\,w^2}}\ ,
\ee
as we shall rediscuss in Sections \ref{sec:axial} and \ref{sec:general}. 

\vspace{2mm}
In such a situation involving axial couplings
we can then view, conversely,  the gauge coupling $g"$ as proportional to $m_U$ and $r$, with
\be
\label{valga}
\dis \frac{g" \cos \xi}{2}\,=\, \dis  \frac{m_U}{v}\ \frac{r}{\sin 2\beta}\,
= \,\dis \,2^{1/4}\,G_F^{1/2} \ m_U \ \frac{r}{\sin 2\beta}\ .
\ee
It also reads 
\be
\label{g"}
\framebox [6.8cm]{\rule[-.35cm]{0cm}{.9cm} $ \dis
\frac{g"}{4}\,\cos\xi \,\simeq \,2\times 10^{-6}\,m_U(\hbox{MeV})\ \frac{r}{\sin 2\beta}\ ,
$}
\ee
which reduces to the simple benchmark value 
\be
\frac{g"}{4}\, \simeq \,2\times 10^{-6}\,m_U(\hbox{MeV})\ r\,
\ee
for $\tan\beta=1$, disregarding for simplicity  $\cos\xi\simeq 1$ \cite{U}.
This is of interest, in particular, when discussing non-standard neutrino interactions with $g(\nu_L)$ 
and $g(e)\!\approx g"/4$, as done in subsections \ref{subsec:mec}, \ref{subsec:mec2}.
As a result  of (\ref{g"}) the axial couplings (\ref{ga00}) read, in the small $m_U$ limit,
 \be
\label{ga0num}
g_{A\pm} 
= \underbrace{\ 2^{1/4}\,G_F^{1/2}\ \frac{m_U}{2}\ }_{\hbox{$2\times 10^{-6}\, m_U(\hbox{MeV})$}}\!\!\!
\times\ 
\left\{\ba{c} r\,\cot\beta\ , \vspace{1mm}\\ r\,\tan\beta\  . \ea\right. 
\ee

\vspace{1mm}

We can also consider a SM-like doublet $h_{\rm sm}$ (possibly replaced by $h_2$ and $h_1^c$, taken with the same quantum numbers so that $\gamma_A=0$), and a semi-inert one $h'$, choosing for convenience, without restriction, the normalisation of $g"\!$ so that 
$h'$ has $F=F'\!=2\,Y\!=\!2$ with $\gamma_{F'}\!=1\,$. 
\linebreak
$F$ is here taken as in  eq.\,(\ref{F-1}) with $\gamma_A=0$.
Then $\eta=2\, v'^2/v^2\!=2\, \sin^2\beta'\,$. 
With $h_{\rm sm}$ and $h'$ having $F=\gamma_Y$ and $\gamma_Y \!+\!2\,$ differing by the same $|\Delta F|=2$ as earlier for  $h_1^c$ and $h_2$,  $m_U$  is still given by the same eq.\,(\ref{mumu})  as in \cite{plb86} with $\beta$ replaced by $\beta'$, independently of $\gamma_Y,\gamma_B$ and $\gamma_{L_i}$. The axial couplings in (\ref{gaeb20}), now isovector, read
\be
\label{valgabis}
g_{A\pm}=\,\mp\ \frac{g"\cos\xi}{2}\ \sin^2\beta'\simeq \ 2^{1/4}\,G_F^{1/2}\ \frac{m_U}{2}\ (\mp\  r\,\tan\beta')\,.
\ee

\vspace{1.5mm}

In general axial couplings may be parametrized 
\vspace{-.5mm}
proportionally to the benchmark value
$\,1/v= $ $2^{1/4}\,G_F^{1/2}\simeq 4 \times 10^{-6}$ MeV$^{-1}$, according to
\be
\label{gaa0}
\framebox[8.6cm]{\rule[-.3cm]{0cm}{.8cm} $ \dis
g_{A\pm}\,=\,2^{1/4}\ G_F^{1/2}\ \frac{m_U}{2}\,A_\pm\,\simeq\, 2\times 10^{-6}\, m_U(\hbox{MeV})\,A_\pm\,.
$}
\ee
$A_\pm$, proportional to the invisibility parameter $r=\cos\theta_A$, is given for the two active doublets 
$h_1,\,h_2$ transforming under $U(1)_A$, by
\be
A_+ =r\,\cot\beta\,,\ \ \ A_-=\,r\,\tan\beta\,,
\ee
and for a SM-like doublet $h_{\rm sm}$ (or $h_2,h_1^c$ with the same quantum numbers) with a semi-inert one $h'$, by
\be
A_\pm\,=\,\mp\ r\,\tan\beta'\, .
\ee

\vspace{-4mm}

\subsection{\boldmath \hspace{-1.5mm}From axial couplings $g_A$ to pseudoscalar ones $g_P\,$}

\vspace{-2mm}

A longitudinal $U$ behaves, in the small mass limit, much as an axionlike particle with effective  pseudoscalar couplings to quarks and leptons obtained from  (\ref{gpga},\ref{gaeb20},\ref{gaa0}) as
\be
\label{gpseudo}
\ba{ccc}
g_{P\,ql}\,=\,\dis \,g_{A\,ql}\ \frac{2m_{ql}}{m_U}\!&=&\!\dis \dis \frac{g"\cos\xi}{2m_U}\ \,m_{ql}\ (\gamma_A\,\mp\,\eta)
\vspace{2mm}\\
\!&=&\!\ \ \underbrace{\  2^{1/4} \, G_F^{1/2}\ \,m_{ql}\  }_{\hbox{$4\times 10^{-6}\, m_{ql}$(MeV)}}\!\!\!\!A_\pm \ .\ \ \ 
\ea
\ee
We get from $g_A$ in (\ref{ga00},\ref{ga0num})
 the effective pseudoscalar couplings of a longitudinal $U$ to quarks and leptons,
expressed in the small $m_U$ limit as 
\be
\label{gpseudo0}
\!g_{P\,ql}= \,g_{A\,ql}\ \frac{2m_{ql}}{m_U}\, = \, \left\{\, \ba{l}2^{1/4} \, G_F^{1/2}\ \,m_u\ r\,\cot \beta \,,
\vspace{2mm}\\
2^{1/4} \,  G_F^{1/2}\ m_{de}\, r\,\tan \beta \,,
\ea
\right.
\ee
as for a quasi-invisible axion;
or similarly, with a semi-inert doublet $h'$,
\vspace{-1mm}
\be
\label{gpseudoh}
g_{P\,ql}
\, = \, \mp\ 2^{1/4} \,  G_F^{1/2}\ m_{ql}\, \ r\,\tan \beta'\,,
\ee 
which vanish in the limit $v'\to 0$\,, for which the semi-inert doublet $h'$ becomes inert.

\vspace{-1mm}

\subsection{\boldmath Recovering the \vspace{1.5mm}pseudoscalar couplings \hbox{of a longitudinal $U$}}
\vspace{-2mm}

These effective pseudoscalar couplings $g_P$ of a longitudinal $U$ boson, proportional to the invisibility parameter $r=\cos\theta_A$, may also be obtained directly from the couplings of the equi\-valent pseudoscalar $a$,   as found from its expression (\ref{rtheta}).
Indeed in the case of (\ref{gpseudo0})
\vspace{-.4mm}
 the Goldstone field 
eliminated by the $Z$ is $\,z_g=\sqrt 2\ \hbox{Im}\,(-\cos\beta\,h_1^0 +\sin\beta \,h_2^0)$, \,and the pseudoscalar $a$ 
reads
\be
\label{expa0}
a\,=\,\sqrt 2 \ \,\hbox{Im} \ [\,\cos\theta_A  \,(\sin\beta \,h_1^0 +\cos\beta\,h_2^0) \,+\, \sin\theta_A\, \sigma \,]\,.\!
\ee
With the Yukawa couplings 
of $h_1^0$ and $h_2^0$ 
\vspace{-.4mm}
expressed as $\lambda_{de}=m_{de}\sqrt 2/(v \cos\beta)$ 
\vspace{-.4mm}
and $\lambda_u=m_u \sqrt 2/(v \sin\beta)$, 
$\sqrt 2\  \hbox{Im} \,h_1^0$ and $\sqrt 2\  \hbox{Im} \,h_2^0$ have 
pseudoscalar couplings to down quarks and charged leptons, and up quarks, 
$m_{de}/(v \cos\beta)$ and $m_u /(v \sin\beta)$, respectively. We thus recover directly the couplings 
 (\ref{gpseudo0}) as 
\be
\label{gugd40}
\left\{
\ba{ccccl}
\dis g_{P\,u}\!&=&\!\dis  \frac{m_u}{v}\ \cos\theta_A\,\cot\beta\!&=&\!\dis  2^{1/4}\,G_F^{1/2}\ m_u\ r\,\cot\beta\,,
\vspace{2mm}\\
\dis g_{P\,de}\!&=&\!\dis  \frac{m_{de}}{v}\,\cos\theta_A\,\tan\beta\!&=&\!\dis  2^{1/4}\,G_F^{1/2}\, m_{de}\ r\,\tan\beta\,.
\ea
\right.
\ee
They are the same as for a standard axion \cite{w1,w2}, multiplied by the invisibility parameter $r$, as for a quasi-invisible axion \cite{U,z,dfs} (see also subsection \ref{subsec:axionlike}).
\vspace{2mm}

Similarly in the case of a semi-inert doublet $h'$ leading to (\ref{gpseudoh}),
\vspace{-.4mm}
the Goldstone field eliminated by the $Z$ is $\,z_g=\sqrt 2\ \hbox{Im}\ (\cos\beta'\,  h_{\rm sm}^0 + \sin\beta'\, h'^0)$, 
so that 
\vspace{.5mm}
\be
\label{ug0}
a=\sqrt 2\ \hbox{Im}\,[\,\cos\theta_A\, (-\sin\beta'\,h_{\rm sm}^0\! +\cos\beta'\,h'^0)+\sin\theta_A \,\sigma\,]\,.
\ee

\vspace{2mm}

\noindent
The Yukawa couplings 
\vspace{-.4mm}
of the SM-like doublet ${h_{\rm sm}^0}$, 
with v.e.v. $v \cos\beta'/\sqrt2$, are enhanced by $1/\cos\beta'$ as compared to the SM.
$\,h_{\rm sm}^0$ (resp.~$h_{\rm sm}^{0*}$, with opposite imaginary part)  plays the role of $h_2^0$ (resp.~$h_1^0$) in giving masses to up quarks (resp.~down quarks and charged leptons).
Multiplying by $\,-\cos\theta_A$ $\sin\beta'\,$ from (\ref{ug0}) and including the extra $-$ sign for down quarks and charged leptons we recover the pseudoscalar couplings of $a$ in (\ref{gpseudoh}) as
\be
\label{gpseudoi}
g_{P\,ql}=\,\mp\ \frac{m_{ql}}{v} \,\cos\theta_A\,\tan\beta'=\mp\ 2^{1/4}\ G_F^{1/2}\ m_{ql}\ r\ \tan\beta'.
\ee

\vspace{-2mm}

\subsection{\boldmath $U$ lifetime, and beam dump experiments}
\label{subsec:tau}

\vspace{-1.5mm}
Let us give the partial widths for $U$ decays into quarks and leptons, from their vector and axial couplings $\,g_{V}=g"\cos\xi\ (Q_U)_V$ and $\,g_A=g"\cos\xi\ (Q_U)_A$ obtained from (\ref{quv}) (see also later (\ref{coup30}-\ref{coup401})).  
From the decay widths into massless particles $\,\Gamma_{f \bar f}= (m_U/12\pi)\,(g_{V\!f}^2+g_{Af}^2)$, 
  \vspace{-.5mm}
and including the phase space factors $(3\beta-\beta^3) /2\,$ and $\beta^3$ for vector and axial couplings, 
 respectively,  
we have with  $\beta_f=v_f/c=(1-4m_f^2/m_U^2)^{1/2}$, $ (3-\beta_f^2)/2=1+2m_f^2/m_U^2\,$, 
\be
\ba{c}
\ba{l}
\hspace{-1.5mm}\Gamma(U\to f\bar f)
\vspace{1.5mm}\\
\!\!=\ \dis \frac{m_U}{12\pi}\ \,\sqrt{1-\frac{4m_f^2}{m_U^2}}\  \left[\,g_{V\!f}^2\,(1+\frac{2m_f^2}{m_U^2} )+g_{Af}^2\,(1-\frac{4m_f^2}{m_U^2} )\right]\ \ 
\vspace{3mm}\\
\!\!= \ \dis \frac{m_U}{24\pi}\ \,\sqrt{1-\frac{4m_f^2}{m_U^2}}\ \times 
\vspace{1.5mm}\\
\hspace{15mm} \dis\left[\,(g_{Lf}^2+g_{Rf}^2)\,(1-\frac{m_f^2}{m_U^2} )\,+\,g_{Lf}\,g_{Rf}\ \frac{6m_f^2}{m_U^2}\right]\,.
\ea
\vspace{-19mm}\\
\ea
\ee

\vspace{19mm}
The resulting lifetime varies considerably depending principally on $m_U$ and $g"$, especially with $g"$ proportional to $m_U$ as given by eq.\,(\ref{g"}) in a 2-doublet\,+\,1-singlet model, leading to 
\vspace{-3mm}
\be
\Gamma_U\,\propto\, g"^2\,m_U \propto \,G_F \,m_U^3\ \frac{r^2}{\sin^2 2\beta}\ .
\ee
This leads in particular, sufficiently above threshold and depending on the axial and/or vector couplings  considered, and accessible channels,  to typical estimates like 
\be
\tau\,
\approx\,\frac{(10^{-9}\ {\rm to} \ 10^{-8})\ {\rm s}}{m_U(\hbox{MeV})^{3}\ r^2}\,
\ee
(reexpressed as $\tau\approx 10^2 \, {\rm y}\,/\,(m_U(\hbox{eV})^{3}\, r^2)$ for a very light $U$ decaying into neutrinos).
The resulting decay length,
\be
l\,=\,\beta_U\gamma_U\, c\,\tau\, \propto \, \frac{p_U}{G_F \,m_U^{4}}\ \frac{\sin^2 2\beta}{r^{2}} \,,
\ee 
is of the order of \cite{U}
\be
l\,\approx\, {.7\ \rm m} \times \ \frac{p_U(\hbox{MeV})}{m_U(\hbox{MeV})^{4}\ r^2}\ ,
\ee
for an axially coupled boson decaying into $e^+e^-$ or $\nu\bar \nu$ pairs, e.g.  $l\simeq 6$ m   for a 20 GeV boson of mass $m_U=7$ MeV with $r=1$.

\vspace{2mm}

The decay length $l$ plays an essential role in the derivation of experimental constraints in terms of $m_U $ and $g"$ (or $r/\sin2\beta$), in particular from beam dump experiments. They require $l$ to be sufficiently short, or sufficiently large,  
for the $U$ boson, if sufficiently produced, to remain undetected. This originally allowed to exclude, from the results of early beam dump experiments at Brookhaven \cite{jacques,coteus} an axially-coupled $U$ boson in the 1 to 7 MeV mass range, in the simplest case $r=\tan\beta=1$ \cite{U}. 
Beam dump experiments can now constrain the existence of such particles in the 1\,--\,10 or up to 1\,--\,100 MeV mass range, depending on the couplings considered and the strength of their interactions
\cite{rev1,rev2}.

\vspace{3mm}
We now present in Secs.\,\ref{sec:cons} and \ref{sec:nu} the main consequences of this analysis for $U$ boson phenomenology.
We leave the more technical aspects on the choice of the extra-$U(1)$ symmetry to be gauged, and on the generation of the $U$ mass and mixing angle $\xi$, for the subsequent  Sections \ref{sec:u1}-\ref{sec:general}.
We finally discuss in  Sec.\,\ref{sec:gut} how the $U$ can find a natural place within the grand-unification (and supersymmetry) frameworks, and resulting implications for its couplings.

\section{\boldmath \vspace{1.6mm}Effects of axial couplings\hbox{:\ }
\hbox{\ $\psi,\Upsilon, \,K,\,B$ \,decays, \,parity violation, ... }}
\label{sec:cons}

\vspace{-.5mm}

\subsection{Quark and lepton couplings}

\vspace{-1.5mm}

Let us write the couplings of chiral quark and lepton fields, as obtained from (\ref{coup0}),

\pagebreak

\vspace*{-12mm}

\be
\label{coup30}
\ba{c}
\ \left\{\ \ba{c}
\dis g_L\,=\,\,g"\cos\xi 
\,\left[\,(\gamma_Y\cos^2\theta -\,\eta\,\sin^2\theta)\ Q \phantom{\frac{\eta}{2}}\,\right.\hspace{10mm}
\vspace{1.5mm}\\
\hspace{25mm}\dis +\left. \frac{\gamma_B}{2}\, B\,+\frac{\gamma_{L_i}}{2}\, L_i  - \,\frac{\gamma_A}{4}\,\pm\,\frac{\eta}{2}\,\right]\,,
\vspace{2mm}\\
\dis g_R\,=\,\,g"\cos\xi 
\,\left[\,(\gamma_Y\cos^2\theta -\,\eta\,\sin^2\theta)\ Q \phantom{\frac{\eta}{2}}\,\right.\hspace{10mm}
\vspace{1.5mm}\\
\hspace{16.5mm}\dis +\left. \frac{\gamma_B}{2}\, B\,+\frac{\gamma_{L_i}}{2} L_i  +\,\frac{\gamma_A}{4}\,\right]\,,
\ea
\right.
\vspace{-10mm}\\
\ea
\ee
\vspace{6mm}

\noindent
valid in the small $m_U$ limit.
The vector and axial couplings $g_V\!=(g_L+g_R)/2$ and $g_A=$ $(g_R-g_L)/2$, defined according to
 \vspace{1.5mm}
\be
\label{avlr0}
g_L\,\gamma^\mu\,\hbox{\small$\dis \frac{1-\gamma_5}{2}$}+g_R\,\gamma^\mu\,\hbox{\small$\dis \frac{1+\gamma_5}{2}$}\,=\,
 \gamma^\mu\,(g_V+g_A\,\gamma_5)\,,
\ee

\vspace{1.5mm}

\noindent
are
\be
\label{coup400}
\ba{c}
\left\{ \ba{ccl}
\dis g_V\!\!&\simeq&\,g"\cos\xi\  \dis
 \left[\phantom{\sum}\hspace{-6mm}\right.\,\gamma_Y\cos^2\theta \,Q +
\frac{\gamma_B}{2}\, B\,+\frac{\gamma_{L_i}}{2}\, L_i 
\vspace{2mm}\\
&&
\ \ \ \ \ \ \ \ \  +\dis  \ \eta \ \hbox{\Large$\left(\right.$}(\,\frac{1}{2}-\,\sin^2\!\theta\,)\,Q-\,\frac{1}{4}\,(B-L) \hbox{\Large$\left.\right)$}
\hspace{-5.5mm}
\left.\phantom{\sum}\right],
\vspace{3mm}\\
\dis g_{A\pm}\!&\simeq&\dis \,g"\cos\xi
\ \ \frac{\gamma_A \dis\mp\,\eta}{4}\,,
\ea
\right.
\vspace{-5.8mm}\\  \ea
\ee

\vspace{6mm}

\noindent
as also seen from (\ref{quv}). The vector couplings may also be reexpressed using
$Q-\,\frac{1}{2}\,(B-L)\,=\,T_3=\pm\,\frac{1}{2}\,$, as
\be
\label{coup401}
\ba{ccl}
\dis g_V\!&\simeq&g"\cos\xi \dis \
 \left[\phantom{\sum}\hspace{-6mm}\right.\,\gamma_Y\cos^2\theta \,Q +
\frac{\gamma_B}{2}\, B\,+\frac{\gamma_{L_i}}{2}\, L_i 
\vspace{2mm}\\
&&
\hspace{20mm}+\dis  \ \eta \ (\,\pm\ \frac{1}{4}-\,\sin^2\!\theta\ Q\,)\hspace{-5mm}\left.\phantom{\sum}\right] \,.
\ea 
\ee

\vspace{2mm}
The seven couplings of chiral quark and lepton fields within a single generation are expressed in (\ref{coup0}) as a linear combination of the five charge operators $Q,\,B,\,L,
\,T_{3L}$ and $F_A$. 
They verify two independent relations:
\be
\label{rel0}
\left\{\,\ba{ccc}
g_{Ae}=\,g_{Ad}\,,
\vspace{1mm}\\
g(\nu_L)=\,g(u_L) +\, g_V(e)-g_V(d)\ .
\ea\right.
\ee
The second one
is related to the fact that the vector part \linebreak in the $U$ current is a combination of $Q,\,B$ and $L$ currents. 
For purely vector couplings to $Q,\,B$ and $L$  it reduces to $\,g(\nu_L)\!=$ $g_V(u) +\, g_V(e)-g_V(d)\,$.
It extends for three generations into
\be
\left\{\,\ba{ccc}
g_{A\,e\mu\tau}=\,g_{A\,dsb}\,,
\vspace{1mm}\\
g(\nu_{iL})-g_V(l_i) = \,g(u_L,c_L,t_L) -g_V(d,s,b)\,.
\ea\right.
\ee

\vspace{1mm}
\subsection{\boldmath Limits on $g_{A\pm}$'s and $g_P$'s \vspace{1.5mm}from $\psi$ and $\Upsilon$ decays}

\vspace{-1mm}

Upper limits on axial couplings $g_A$ and associated pseudoscalar ones $g_P$ from $\psi$ and  $\Upsilon$ decays 
are rather stringent.
With the correspondence $\,r\,\cot\beta\to  A_+$,\  $r\,\tan\beta $ $\to $ $A_-$,  we get the branching ratios \cite{U,prd07}, 
\be
\label{psi2}
\hbox{$
\left\{\!\ba{ccccc}
\dis \frac{B\,(\,\psi  \to \gamma \, U\,)}{B(\psi\to \mu^+\!\mu^-)}\!\!&\simeq&\!\dis
\frac{G_F m_c^2}{\sqrt 2\,\pi\alpha}\ C_\psi A_+^2\!\!&\simeq&\!  8\times 10^{-4}\, C_\psi A_+^2,
\vspace{2mm}\\
\dis \frac{B\,(\,\Upsilon  \to \gamma \, U\,)}{B(\Upsilon\to \mu^+\!\mu^-)}\!\!&\simeq& \!\dis
\frac{G_F m_b^2}{\sqrt 2\,\pi\alpha}\ C_\Upsilon A_-^2\!\!&\simeq&\!  8\times 10^{-3} \, C_\Upsilon A_-^2,
\ea
\right.
$}
\ee
so that 
\be
\left\{\ \ba{ccc}
B(\psi\to\gamma \,U)\, \simeq \,5 \times 10^{-5}\ C_\psi\,A_+^2 \,,
\vspace{2mm}\\
B(\Upsilon\to\gamma \,U)\, \simeq \,2 \times 10^{-4}\ C_\Upsilon\,A_-^2 \,.
\ea
\right.
\ee
$C_\psi$ and $C_\Upsilon$, expected to be larger than 1/2, take into account QCD radiative and relativistic corrections.
We obtained long ago 
\be
\left\{\ 
\ba{ccc}
|\,g_{A\,uct}\,| \!&\simle&\!1.5\times10^{-6}\ m_U(\hbox{MeV})/\sqrt{B_{\rm inv}}\,,
\vspace{2mm}\\
|\,g_{A\,e\mu\tau}=g_{A\,dsb}\,| \!&\simle&\!.8\times 10^{-6}\ m_U(\hbox{MeV})/\sqrt{B_{\rm inv}}\,.
\ea\right.
\ee

\vspace{3mm}

The improved experimental limits for a light $U$ \cite{psi1,psi2,ups1,ups2,ups3},
\be
\left\{\ \ba{ccr}
B(\psi\to\gamma + \hbox{invisible} \ U)& < &7\times 10^{-7}\,,
\vspace{2mm}\\
B(\Upsilon\to\gamma + \hbox{invisible} \ U)& \simle &\ \ 10^{-6}\,,
\ea
\right.
\ee
now  imply 
\be
\label{limcop}
\left\{\ \ba{ccc}
|A_+| &<& .17  /\sqrt{B_{\rm inv.}}\ ,
\vspace{2mm}\\
|A_- |&\simle& \  .1 /\sqrt{B_{\rm inv.}}\ ,
\ea\right.
\ee
corresponding, as seen from (\ref{ga0num},\ref{gaa0}), to
\be
\label{limgab0copie}
\left\{\ \ba{ccr}
|g_{A\,c}|&\simle & 3.4 \times 10^{-7}\ m_U(\hbox{MeV})/ \sqrt{B_{\rm inv}}\,,
\vspace{2mm}\\
|g_{A\,b}| & \simle &\ 2 \times 10^{-7}\ m_U(\hbox{MeV})/ \sqrt{B_{\rm inv}}\,.
\ea\right.
\ee
They may be combined into
\be
\sqrt{\,|A_+\,A_-|}\ \simle\   (.13 \ \,\hbox{to}\  .16) /\sqrt {B_{\rm inv}}\,,
\ee
for $m_U$ in the 0 to 1 GeV/$c^2$ mass range.
When $A_+=r\,\cot \beta$ and $A_-= r\,\tan\beta$ this reads
 \be
r\ \simle\   (.13 \ \,\hbox{to}\  .16) /\sqrt {B_{\rm inv}}\,,
\ee
 independently of $\tan\beta$.
A longitudinal $U$ boson is then required to behave mostly as a spin-0 electroweak singlet, as anticipated long ago.

\vspace{2mm}

When the axial part in the $U$ current originates totally, either from the universal axial current $J^\mu_A$, isoscalar,
or from the mixing with the $Z$, then
$|g_{A\,uct}|=|g_{A\,e\mu\tau}|=| g_{A\,dsb}|$ as in (\ref{univuniv}).
The Yukawa couplings of  $a$ in (\ref{gpga}) are all proportional to quark and lepton masses up to a possible sign, with 
 \be
 \ba{ccl}
 \dis |g_{P\,ql}|= \frac{m_{ql}}{v}\ A\!&\simeq&\ \ 2^{1/4}\, G_F^{1/2}\, m_{ql}\ A
 \vspace{2mm}\\
 \!&\simeq&\!4\times 10^{-6}\,m_{ql}(\hbox{MeV})\ A\,.
 \ea
 \ee
$\psi$ and $\Upsilon$ decays then cease to give complementary indications on the 
couplings to $c$ and $b$ quarks. 
It is thus important that $\psi$ decay results \cite{psi2} be expressed independently of $\Upsilon$ results \cite{ups3}, to be applicable in a model-independent way (not just with NMSSM-like situations in mind). The present limit from $\Upsilon $ decays then implies
\be
|A_+|=|A_-|\,\simle\,.1/\sqrt{B_{\rm inv}}\,.
\ee
\vspace{-3mm}

This applies in particular to semi-inert doublet models with isovector axial couplings,  for which $A^\pm=\mp\, r\,\tan\beta'$ as in (\ref{gpseudoi}). Then a small $A$ does not necessarily imply a small $r$ as $\tan\beta'$ may also be small. One might even have $r=1$ i.e. no extra singlet, with a small $\beta'$ i.e. a small semi-inert doublet v.e.v. $v'/\sqrt 2\,$ (as compared with $v/\sqrt 2\simeq 174$ GeV).  
This is not surprising as for $v'\to 0\,$ the semi-inert doublet model tends towards an inert-doublet one with $a$ in (\ref{ug0}) becoming decoupled from quarks and leptons, even with no extra singlet. Still such a  possibility (not strongly motivated)  gets disfavored in view of the stronger limits on 
$|A_+|=|A_-|\,$ from $K$ and $B$ decays (see later eq.\,(\ref{lima+}) in subsection \ref{subsec:other}), then requiring $v'$ to be much smaller than $v$.

\vspace{-2mm}

\subsection{\boldmath Axionlike couplings to $e$'s must be \hbox{very small}}

\vspace{-1mm}

Let us move to the discussion of effective pseudoscalar axionlike couplings of the $U$ boson to electrons.
The universality relations (\ref{univ}) imply
 that $g_{Ae}$ and $g_{A\mu}$ must obey the same bounds as obtained for $g_{Ab}$ from $\Upsilon$ decays \cite{pl09}, or for $g_{As}$ from $K^+$ decays, thus, with $|A_-|\simle .1/\sqrt{B_{\rm inv}}$ as deduced from \cite{ups3} ,
 \be
\label{limgae0}
\framebox [7cm]{\rule[-.7cm]{0cm}{1.6cm} $ \dis
\left\{\ \ba{ccl}
|g_{Ae}|\!&\simle&\, 2 \times 10^{-7}\ m_U(\hbox{MeV})/ \sqrt{B_{\rm inv}}\,,
\vspace{3mm}\\
 |g_{Pe}| \!&\simle & \,2 \times 10^{-7} / \sqrt{B_{\rm inv}}\,.
 \ea \right.
 $}
 \ee 
 $B_{\rm inv}$ denotes the branching ratio for the $U$ to be undetected in the experiment considered, 
due to a long enough lifetime or to invisible decay modes into $\nu\bar\nu$ or light dark matter particles.
 For small $\sqrt{B_{\rm inv}}$ complementary results can be obtained from visible $U$ decays into
 $e^+e^-$ or $\,\mu^+\!\mu^-$, depending on $m_U$.
 
 \vspace{2mm}

 \vspace{2mm}
 Furthermore, if the axial couplings are purely isoscalar or isovector so that
 $|g_{Ae}|=|g_{At}|$, the even stronger constraint that may be derived on the latter (see later eqs.\,(\ref{limgat},\ref{limgat2}) in subsection \ref{subsec:other}) can also be applied to the electron, then  leading to
 \be
 \label{gaest}
 \left\{\ 
 \ba{ccl}
  |g_{Ae}|&\simle&\ 2 \times 10^{-9}\ m_U(\hbox{MeV})\,,
  \vspace{2mm}\\
    |g_{Pe}|&\simle&\ 2 \times 10^{-9}\ .
    \ea \right.
  \ee

\subsection{\boldmath An extra $U(1)$ broken at the $\,\approx 10^8$ GeV scale ?}
\label{subsec:xe}

\vspace{-1mm}

What about even smaller values of $g_{Ae}$ and $g_{Pe}$ ?  \,A very light $U$ boson with energies in the keV range and very small effective pseudoscalar coupling to electrons $g_{Pe}\approx 3\times 10^{-12}$ might possibly be responsible for the excess electronic recoil events recently observed in  XENON1T, 
if attributed to a solar axionlike particle rather than to some other effect \cite{xenon}.
Such a small $g_{Pe}$, and associated axial coupling
\be
g_{Ae}\simeq \, \frac{m_U}{2m_e}\ g_{Pe}\,\approx\, 3\times 10^{-15} \ m_U(\hbox{keV})\,,
\ee
would then correspond, using the expression of the effective pseudoscalar coupling to the electron as
\be
g_{Pe}\,\simeq\,2\times 10^{-6} \ r \,\times\,  (\tan\beta \ \,\hbox{or}\  \tan\beta')\,,
\ee 
from eqs.\,({\ref{ga0num},\ref{gaa0},\ref{gpseudo}) or (\ref{gpseudoi}), to 
\be
\label{a6}
A_-=\,r\ (\tan\beta \ \, {\rm or}\  \tan\beta')\ \approx\, 1.5\times 10^{-6}\ ,
\ee
so that $g_{Pe}\simeq \,2\times 10^{-6}\,A_-\!\approx 3\times 10^{-12}$\ (leaving aside possible astrophysical constraints).

\vspace{2mm}

Such a small value of $|A^-|$  may be associated with a very small value of the invisibility parameter $r$ \cite{U} originating from a large singlet v.e.v. typically $\approx 10^6$ times the electroweak scale. More precisely this large singlet v.e.v. $w/\sqrt 2\,$, obtained from (\ref{r20}) as
\be
\ba{ccl}
|F_\sigma |\,w&\simeq &\dis\frac{v\,\sin2\beta}{r}\,\simeq\, \frac{v\ 2\sin^2\!\beta}{A_-}
\vspace{2mm}\\
&\simeq &\dis 1.6 \times 10^8\ \hbox{GeV}\,\times \,(2\sin^2\!\beta)\,,
\ea
\ee
should then be of the order of $\,\approx 10^8$ GeV.
(In the absence of a singlet i.e. with $r=1$ one may also consider a very small semi-inert v.e.v. 
$v' \approx 1.5\times 10^{-6} \,v\simeq 370$ keV  as seen from (\ref{a6}), but this would require a large amount of fine tuning.)
 
 \vspace{-1mm}
 
 \subsection{Parity-violation effects in atomic physics}
 
 \vspace{-1mm}
 
A positive aspect of a very strong constraint on $g_{Ae}$, as in (\ref{gaest}),
 is that it facilitates considerably satisfying another constraint from parity-violation effects in atomic physics. This one requires \cite{bf}, for the change in the weak charge $Q_W$ of the caesium nucleus to be less than 1 \cite{pdg},
\be
\label{pvat}
\sqrt{\, |\,g_{Ae}\,g_{V\!q}\,|}\ \simle \,10^{-7}\ m_U(\hbox{MeV})\,,
\ee
corresponding to
\be
\frac{|\,g_{Ae}\, g_{V\!q}\,|}{m_U^2}\simle\, 10^{-3}\ G_F\ ,
\ee
 for $m_U$ larger than a few MeV/$c^2$. 
 
 \vspace{2mm}
 
More precisely the axial coupling of the electron $g_{Ae}$ and the vector coupling  $g_{V\!q}$ of an average quark in the nucleus are obtained from (\ref{quv},\ref{gaeb20}) as
 \be
 \label{pvat2}
 \left\{
 \ba{ccl}
 \ \ g_{Ae}\!&=&\! \dis g"\cos\xi\ \frac{\gamma_A+\eta}{4}\,,
 \vspace{2mm}\\
3\,(Z\!+\!N)\,g_{V\!q}\!&=&\! \dis g"\cos\xi\ \hbox{\Large$\left[\right.$}  \,\gamma_Y\cos^2\theta \ Z +\dis\frac{\gamma_B}{2}\ (Z\!+\!N)
\vspace{1.5mm}\\
&& \!\dis \ \ \ \ \ \ +\ \eta\ \hbox{\large$\left[\right.$} (\,\frac{1}{4}-\,\sin^2\theta\,)\ Z -\,\frac{1}{4}\, N\hbox{\large$\left.\right]$}\,
\hbox{\Large$\left.\right]$}\  ,
 \ea \right.
 \ee
 
 \vspace{1mm}
  \noindent
 allowing to express $\Delta Q_W$, in present notations, as \cite{bf},
 \be
 \label{pvat3}
 \framebox[7.45cm]{\rule[-.45cm]{0cm}{1.15cm} $ \dis
 \Delta Q_W\,=\,\frac{2\sqrt 2}{G_F}\ \frac{g_{Ae}\!\times \,3\,(Z\!+\!N)\,g_{V\!q} }{m_U^2 }\ K(m_U)\, .
 $}
 \ee
 $K(m_U)$ takes into account the effect of the $U$ mass, varying between about .5\, for $m_U$ of a few MeV, approaching 1 for $m_U$ larger than about 50 MeV.
 
 \vspace{2mm}
 
Let us check this general formula on the very specific case of a dark $Z$, the comparison with the $Z$ exchange amplitude being immediate in this case. Ignoring for simplicity $\cos\xi\simeq 1$,  we have in this approximation, with $\gamma_Y\!=\gamma_B\!=\gamma_{L_i}\!=\gamma_A\!=0,\ \eta=2\sin^2\beta'$, 
 \be
 g_{Ae}\!= \frac{g"}{4}\,\eta ,\ \ 3\,(Z\!+\!N)\,g_{V\!q}= \frac{g"}{4}\,[(1-4\sin^2\theta)\,Z-N]\,\eta\, ,
 \ee
 so that
 \be
 \Delta Q_W=\dis \frac{m_Z^2}{g_Z^2}\ \frac{g"^2}{m_U^2} \ \eta^2\ [(1-4\sin^2\theta)\,Z-N]\ K(m_U)\,.
 \ee
It is proportional to $Q_W(Z,N)$, and may be reexpressed 
with $\tan\xi=(g"/g_Z)\,\eta$ from (\ref{txieta}), as
\be
 \Delta Q_W=\, \tan^2\xi\ \frac{m_Z^2}{m_U^2}\ Q_W(Z,N)\ K(m_U)\,,
\ee
as immediately found for a dark $Z$ of mass $m_U$ rather than $m_Z$, coupled to the $Z$ current with relative strength $\tan\xi$  \cite{mar,marbis}.
With 
\be
\label{xidz}
\tan\xi\ \,\frac{m_Z}{m_U}=\ \frac{g"}{g_Z}\, \frac{m_Z}{m_U}\ \eta\,=\, \frac{g"v}{2m_U}\ \eta
=\frac{r}{\sin2\beta'}\, \eta\,=r\,\tan\beta'
\ee
from (\ref{xi0},\ref{delta0}) and (\ref{valga},\ref{valgabis}),  the relative effect of a dark $Z$ on $Q_W$ is given by
\be
\frac{\Delta Q_W}{Q_W}\,=\,r^2\,\tan^2\beta'\ K(m_U)\,.
\ee

This effect is $\simle 1\%$ as experimentally required if the invisibility factor $r$ is small enough, typically $\simle .1$; or if the semi-inert v.e.v. $v'/\sqrt 2$ is small enough (typically $\simle 17$ GeV for $r=1$) so that $\tan\beta'$ is small. Or also if $K(m_U)$ is small, with e.g. $K\simle .025$ for $m_U\simle .1 $  MeV, atomic physics experiments loosing some of their sensitivity for a too light $U$ boson. But the factor $A=r\,\tan\beta'$ gets in any case more strongly constrained from $B$ and $K$ decay experiments than from atomic physics parity-violation, in the case of a dark $Z$.
\vspace{2mm}

More generally returning to eqs.\,(\ref{pvat}-\ref{pvat3}) we get in a similar way, for $g_{Ae}\simeq (g"/2)\sin^2\beta'\simeq [m_U/(2v)] \, (r\,\tan\beta')$ 
and $g_{V\!q}\propto g"/2\,\simeq (m_U/v) \ r/\sin2\beta'$,
\be
\frac{\Delta Q_W}{Q_W}\,\propto\,r^2\,\frac{1}{\cos^2\beta'}  \ K(m_U)\,.
\ee
Similarly with the active doublets $h_1,h_2$ leading to   $g_{Ae}$ in (\ref{ga00},\ref{ga0num}), we get, with 
$g_{Ae}\simeq (g"/2)\sin^2\!\beta\simeq [m_U/(2v)]$ $\times \, (r\,\tan\beta)$, 
\be
\frac{\Delta Q_W}{Q_W}\,\propto\,r^2\,\frac{1}{\cos^2\beta}  \ K(m_U)\,.
\ee
The exact expressions depend on $\gamma_Y,\gamma_B$ as well as on $\gamma_A$ and $\eta$,  in particular through $g_{V\!q}$ in (\ref{pvat2}).
\vspace{-2mm}

\subsection{Other limits from meson decays}
\label{subsec:other}

\vspace{-1mm}
A light $U$ boson may be produced in meson decays.
A preliminary estimate of the decay rate for $K^+\to \pi^+\,U$ was proposed in \cite{U} before the top quark was known or even imagined to be so heavy, based on
 \be
\left\{\ \ba{ccl}
\dis g_{As}
\!&=& 2\times 10^{-6}\ m_U(\hbox{MeV})\ A_-\,,
\vspace{1mm}\\
\dis g_{Ps}\!&=&\dis g_{A\,sl}\ \frac{2m_s}{m_U}\,\simeq\, 6\times 10^{-4}\ A_-\,,
\ea
\right.
\ee
and

\vspace{-6mm}

\be
 \label{ku2}
 B(K^+\! \to\pi^+ U)\,\propto\, \frac{1}{\pi}\ \frac{g_{Ps}^{\ 2}}{4\pi}\,\simeq\,10^{-8}\,A_-^2\,,
 \ee
leading to consider an order of magnitude estimate in the $\,\approx 10^{-8}\,A_-^2$ (up to possibly $\approx 10^{-5} \,A_-^2$) range.  This led tentatively, from 
\be
\label{ku}
B(K^+\!\to \pi^+\! +\hbox{invisible}\  U)<(.73\ \,\hbox{to} \approx\,1) \times 10^{-10}\,
\ee
for $m_U< 100$ MeV  \cite{K},  in the absence of cancellation effects involving large couplings to the top,
 to $|A_-|\simle.1\,$ for an invisible $U$, i.e. \cite{prd06}
 \be
 |g_{As}|\,\simle\ 2 \times 10^{-7}\ m_U(\hbox{MeV})/ \sqrt{B_{\rm inv}}\,.
 \ee
This is also what is now found for $|g_{Ab}|$ from $\Upsilon$ decays in (\ref{limgab0copie}).

 \vspace{2mm}
 
 The top quark, however, is very heavy, leading to
\be
g_{Pt} \,= \,\frac{2m_t}{m_U}\ g_{At}\,=\,\frac{m_t}{v}\ A_+\,\simeq \,.7\  A_+\,,
\ee
resulting in 
$({1}/{\pi})$ ${g_{Pt}^{\ 2}}/{4\pi} \simeq 1.2\times 10^{-2}\,A_+^2\,$. 
For processes involving the coupling of a longitudinal $U$ to a $t$ quark the branching ratios may be expressed as
\be
\label{bkb}
\left\{\ 
\ba{ccc}
B(K^+\! \to\pi^+  U)
\vspace{2mm}\\
B(B \to K\,U)
 \ea\right.
\propto \, \frac{1}{\pi}\, \frac{g_{Pt}^{\,2}}{4\pi}\,\simeq \,1.2\times 10^{-2}\, A_+^2\,.
\ee
These ratios have been estimated in  \cite{mar,marbis} in the special case of a dark $Z$, for which $\gamma_Y\!=\gamma_B=\gamma_{L_i}\!=\gamma_A\hspace{-.5mm}=0\,$, \linebreak
$\eta= 2\,\sin^2\!\beta'$,  
\,$\tan\xi \hspace{1.2mm}m_Z/m_U\simeq \,r\,\tan\,\beta'=A_+$ as in  (\ref{xidz}),  then denoted in this specific case by 
$\delta$. The results may be transposed to general situations dealing with the production a light $U$ from the axial top quark current, allowing us to express the rates (\ref{bkb}) in terms of $A_+$, even if this one is in general no longer to be identified with
$\delta = \tan\xi \hspace{1.2mm}m_Z/m_U$ (due to $\gamma_Y$ in $\tan\xi=(g"/g_Z)\ (\gamma_Y+\eta)$ in (\ref{txieta})). And indeed $\tan \xi\neq 0$ for a dark photon, which has no significant axial couplings.

\vspace{2mm}

We can then write
\be
\left\{\ 
\ba{ccc}
B(K^+\to\pi^+ U)\!& \approx &\!4\times 10^{-4}\,A_+^2\,, 
\vspace{2mm}\\
 B(\,B\ \to\,K\,U\,)\!&\approx&\! .1\ A_+^2\,.\ \ \ \ 
\ea
\right.
\ee
This leads to stronger bounds on $|A_+|$, such as, from kaon decays and for $m_U\simle $ 100 MeV, 
\be
\label{lima+}
|A_+| \simle 10^{-3}/\sqrt{B_{\rm {inv}}}\,;
\ee
and even down to $\simle 10^{-4}/\sqrt{B_{\rm {inv}}}\,$, depending on $m_U$ \cite{kb}.
Considering also charged decay modes it is reasonable to retain the estimate $|A_+|\simle10^{-3}$, corresponding to
 \be
\label{limgat}
\left\{\ \ba{ccl}
\ \ |g_{At}|\,\simle\, 2 \times 10^{-9}\ m_U(\hbox{MeV})\,,
\vspace{2mm}\\
\hbox{or}\ \ \dis  |g_{Pt}|\,=\, \frac{2m_t}{m_U}\,|g_{At}| \,\simle\,.7\times 10^{-3} \,.
 \ea\right.
\ee
For a purely isoscalar or purely isovector axial coupling this leads to a very small effective pseudoscalar coupling to the electron,
\be
\label{limgat2}
\dis  |g_{Pe}|\,=\, \frac{m_e}{m_t}\ |g_{Pt}| \ \simle\ 2\times 10^{-9} \,,
\ee
a limit still allowing for the much smaller axionlike electron coupling $\simeq \,3\times 10^{-12}$ tentatively discussed in subsection \ref{subsec:xe}.

\section{\boldmath \vspace{1.5mm}Neutrino interactions, \hbox{$\pi^0\to\gamma\,U$ decays}, \,$U$-induced forces, ...}
\label{sec:nu}

\subsection{\boldmath Non-standard neutrino\vspace{1.2mm} interactions, \hbox{\ \ \ \ with a small $\,g"\propto m_U$}}
\label{subsec:mec}

\vspace{-2mm}

The exchanges of a new $U$ boson would lead to non-standard neutral-current neutrino interactions, a prime potential  effect discussed forty years ago \cite{U},  that would not exist for a pure dark photon, uncoupled to neutrinos. They must be somewhat smaller than the standard ones, otherwise they would have been detected already. The comparison between non-standard and standard neutrino interactions involves at first comparing $\,{g"^2}/(8m_U^2)$ 
with $G_F/\sqrt 2= {g_Z^2}/(8m_Z^2)$. It leads to a benchmark value
\be
\frac{g"}{4}\,=\,2^{-3/4}\,G_F^{1/2}\,m_U\,\simeq\,2\times 10^{-6}\,m_U(\hbox{MeV})\,,
\ee
such that 
\vspace{-3mm}
\be
\label{bench2}
\frac{g"^2}{8m_U^2}\,=\,\frac{g_Z^2}{8m_Z^2}\,=\,\frac{G_F}{\sqrt 2}\,,
\ee
with $g"$ naturally proportional to $m_U$.
\,Furthermore in the presence of two doublets and a singlet with an invisibility parameter $r\leq 1$, we have as in  \cite{plb86}
\be
\label{bench}
\frac{g"}{4}\,\cos\xi =\,\frac{m_U}{2v}\,\frac{r}{\sin2\beta}\simeq \,2\times 10^{-6}\, m_U(\hbox{MeV})\ \frac{r}{\sin 2\beta}\ \ 
\ee
 (cf.\,also Sec.\,\ref{sec:axial}). It serves in the evaluations of the $U$ cou\-plings $\,g"\cos\xi \ Q_U$, 
as in  (\ref{ga0num}) for the axial couplings $g_A$.

\vspace{2mm}
The couplings to neutrinos should then generally be small, especially for small values of $m_U$. 
Indeed 
from $\nu_e\,e$ scattering experiments at low energies at LAMPF and LSND \cite{neu1,neu2} we got, some time ago, a limit on the product of couplings 
$g(\nu_L)g(e)\simeq g"^2 Q_U(\nu_L)Q_U(e)$
\cite{ldm1,prd07},
\be
\ba{ccc}
\sqrt{\,|g(\nu_L)\,g(e)|}\ \simle\  3\times 10^{-6}\ m_U(\hbox{MeV})
\vspace{2mm}\\ 
\hspace{10mm}\hbox{for $m_U \simge $ a few MeV's}\,.
\ea
\ee
\vspace{-2mm}

\noindent
This limit, of about $3 \times 10^{-5}$ for $m_U\simeq 10$ MeV, applies in particular for a coupling to $B\!-\!L$\,, which should then also verify 
$|g_{B-L}] <3\times 10^{-6}\ m_U(\hbox{MeV})$.
 It remains relevant \linebreak in present discussions of neutrino-electron scattering experiments, which lead to similar limits, though more constraining for lower values of $m_U$ \cite{neu3,neu4}.

\vspace{-2mm}

\subsection{Three mechanisms\vspace{1.2mm} for small non-standard neutrino interactions}
\label{subsec:mec2}
 
 \vspace{-2mm}
 
More precisely, with two BEH doublets $U$-exchange amplitudes between two particles 1 and 2 are found
(using expression (\ref{valga}) of $g"\cos\xi$) to be proportional to  
\be
\label{amp}
\ba{l}
\dis \frac{g"^2\cos^2\xi}{8\,(m_U^2-q^2)}\, Q_{U1}Q_{U2} = \frac{G_F}{\sqrt 2} \, \frac{m_U^2}{m_U^2-q^2}\ \frac{r^2}{\sin^2 2\beta}\ Q_{U1}Q_{U2}\,,
\vspace{.5mm}\\
\hspace{50mm} \uparrow \hspace{11.5mm}\uparrow\hspace{10.5mm} \uparrow
\vspace{.5mm}\\
\hspace{49mm} \bf1)\hspace{10.5mm} 2)\hspace{9mm} 3)
\ea
\ee

 \noindent
 to be compared with SM amplitudes, 
 $\propto \!G_F/{\sqrt 2}\,$. 
This expression plays an essential role to ensure that these non-standard neutrino interactions be sufficiently small.
Beyond the choice of a small  extra-$U(1)$ gauge coupling, with $g"$ naturally proportional to $m_U$ as seen above, 
having this expression small as compared with $G_F/\sqrt 2$ may be realized using one or the other of three mechanisms (possibly combined) acting on the factors 1), 2) or 3) in (\ref{amp}) \cite {U} :

\vspace{2.5mm}
{\bf 1)} {\it A sufficiently small $m_U$ 
compared to $\sqrt{|q^2|}$ in the experiment considered}, \,which motivated very early  {\it a light or even very light $\,U\!$ boson}, so that $m_U^2/|q^2|$ be small. For the scattering of low-energy neutrinos on electrons at rest, this means, for example, favoring values of $m_U$ smaller than 
$\sqrt{\,2m_e E_{\rm recoil}}\,$. 

\vspace{1.5mm}

In view of the recent interest for interpreting the possible excess of events with electronic recoil  energies of a few keV's observed by XENON1T \cite{xenon}, this would mean, with $E_{\rm recoil} \approx $ 2.5 keV, favoring 
values of $m_U\simle 50$ keV. Such a light $U$  would also make $U$ effects generally too small to be detected in experiments performed at a higher $q^2$,
thanks to the small $m_U^2/(-q^2)$ factor in (\ref{amp}). 

\vspace{1.5mm}

For $m_U\!\approx 50$ keV the central value of the gauge coupling $g"$ in (\ref{bench2},\ref{bench}) is $\approx 4\times 10^{-7}$, so that
\be
\ba{c}
m_U\simle 50 \ \hbox{keV}\ \  \hbox{with} \  \  g"\!\approx 4\times10^{-7}\ \Rightarrow \hspace{10mm}
\vspace{2mm}\\
\hspace{20mm}\dis \frac{g"^2}{8\,(m_U^2\!-q^2)}\approx\ (.5 \ {\rm to}\ 1)\ \, \frac{G_F}{\sqrt 2}\ 
\ea
\ee

\noindent
for $E_{\rm recoil} \approx $ 2.5 keV, allowing for a comparison with standard solar neutrino interaction amplitudes.
Not very surprisingly when dealing with corrections to weak-inter\-action cross-sections, this is what may be needed for a possible interpretation of the excess (should it be due to new physics\,...), depending also on the values of $Q_U(\nu_L)$ and $Q_U(e)$ \cite{x2,x3,x4}.

\vspace{3.5mm}

{\bf 2)} The $U$-induced amplitudes in (\ref{amp}) may be reduced thanks to {\it a small value of the invisibility parameter $r$} associated with a large singlet v.e.v., with $g"$ expressed proportionally to $m_U$ and $r$ as in (\ref{valga}), thereby further decreasing the value of $g"$ for any given $m_U$ as shown by (\ref{bench}).

\vspace{3.5mm}

{\bf 3)} These amplitudes may also be reduced owing to  {\it small or even vanishing values for the
$U\!$ charges of neutrinos}.
This was also initiated in \cite{U}, noting that the $U$ charge of neutrinos in (\ref{coup0}) would vanish for a coupling to $F_A+L/2$
corresponding to 
$\gamma_A\!=1,\gamma_L\!=1/2$, without mixing with $Z$ ($\gamma_Y=\eta=0$).
\,We can also consider couplings to $B,\ Q$ and $L-2\,T_{3L}$, or equivalently $T_{3R}=$ $Q-T_{3L}-\frac{1}{2}(B-L)$, \,which all vanish for $\nu_L$.
In gene\-ral $Q_U$ is expressed   in (\ref{coup0}) as a five-parameter (assuming lepton universality) linear combination of $Q, \,B,\,L,\,T_{3L}$ and $F_A$\,,  so that we have for neutrinos
\be
g(\nu_L)\simeq \,g"\cos\xi\ (\,\frac{\gamma_L}{2} -\,\frac{\gamma_A}{4}+\,\frac{\eta}{2}\,)\,.
\ee
Thus 
\be
\framebox[7.45cm]{\rule[-.55cm]{0cm}{1.3cm} $ \dis
\ba{ccc}
g(\nu_L)= 0 \ \ \ \hbox{$\Leftrightarrow \ Q_U=$ \,linear combination of} 
\vspace{2mm}\\
Q,\ B,\ T_{3R}\,\ \hbox{and}\ \,L+2\,F_A\,.
\ea
$}
\ee
Some of these operators may be replaced by other combinations like $F_A\!+T_{3L}$ or $L-2\,T_{3L}$, ...\,.
If we exclude $F_A$ a vanishing coupling to neutrinos simply corresponds to having $Q_U\!$ as a linear combination of $Q$, $B$ and $T_{3R}$.

\vspace{1.5mm}

An early example of such a situation is obtained for an axial current combined with the $Z$ current as in (\ref{juold2}) \cite{plb86}, with
$\gamma_A=1$, $\eta=-\cos2\beta$, so that
\be
\label{nphob0}
g(\nu_L)\,=\,g"\cos\xi \ (\,-\frac{1}{4}-\,\frac{1}{2}\,\cos 2\beta)\,,
\ee
vanishing for $\cos 2\beta=-1/2$. For $\beta\simeq \pi/3$, $\eta=1/2$,
\be
\label{nphob}
Q_U=\,\frac{1}{2}\, (F_A+T_{3L}-\sin^2\!\theta\,Q)\ \  \hbox{is \it ``neutrinophobic''.}
\ee
The possibility of a small or even vanishing $Q_U$ is something to keep in mind when discussing non-standard neutrino interactions and interpreting possible excesses, especially if it turns out that there are no such excesses.

 \subsection{\boldmath $\pi^0\to\gamma\,U$ decays and possible protophobia}
 
 \vspace{-1mm}
The singlet positronium state (parapositronium) with $C=+$ and decaying into $\gamma\gamma$,  can also decay into $\gamma\,U$ through the vector coupling of the electron but not the axial one, with a branching ratio 
\be
B (^1S_0\ {\rm Ps} \to\gamma\,U)\,\simeq \,2\,\left(\frac{g_{V\!e}}{e}\right)^2\,,
\ee
in the small $m_U$ limit \cite{U}.
In a similar way the $\pi^0$ can decay into $\gamma\,U$ through the vector couplings of the $U$ to $u$ and $d$ quarks, with a branching ratio \cite{do}
\be
B (\pi^0 \to\gamma\,U)\,\simeq \,2\,\left(\frac{2g_{V\!u}+g_{V\!d}}{e}\right)^2= 2\,\left(\frac{g_V(p)}{e}\right)^2\,,
\ee
 Axial couplings do not contribute, the axial quark current being a $C=+$ operator, as indicated in (\ref{cp})

\vspace{2mm}
The $\pi^0\to \gamma\,U$ decay rate
gets suppressed for small $g_V(p)\simeq g"\,(Q_U)_V(p)$.  This may be obtained with a small value of the gauge coupling $g"$, naturally proportional to $m_U$ as we saw. This may also be obtained with a suppressed value of the $U$ charge of the proton, as naturally obtained long ago  \cite{plb86} as a result of $Z$-$U$ mixing effects, with a $U$ charge $(Q_U)_V=\eta \,(Q_Z)_V$ in (\ref{quv},\ref{qzu},\ref{qzu2}), where
\be
\label{qzu8}
(Q_Z)_V= (T_{3L})_V-\sin^2\theta \,Q = (\,\frac{1}{2}-\,\sin^2\theta\,)\ Q-\,\frac{1}{4}\ (B-L)\ .
\ee

\vspace{-2mm}

\noindent
Then
\be
\label{gvprot}
\frac{g_V(p)}{g_V(n)}\,=\,\frac{Q_{\rm Weak}(p)}{Q_{\rm Weak} (n)}\,\simeq \,4\sin^2\theta -1\,\simeq\,-\,0.07\,,
\ee
using the recent  values of $p$ and $n$ weak charges \cite{qw}. The $U$ boson is then naturally ``protophobic''.
Such an inhibition of $\pi^0\!\to\gamma\,U$ decays gets increasingly favored by the recent limits from the 
 NA48/2 experiment \cite{NA48,NA482}, in particular, requiring 
 \be
 \ba{ccc}
|g_V(p)| \!&=&\!g"\cos\xi \ |(Q_U)_V(p)|\,= \,|\epsilon_{V\!p}\,e|
\vspace{2mm}\\
& \simle& \ 3\times 10^{-4}/ \sqrt{B(U\!\to\! e^+e^-)}\ .
\ea
\ee
Eq.\,(\ref{gvprot}) for $g_V(p)$  is valid when the vector part in the $U$ current mostly originates from the mixing with the $Z$ as in (\ref{juold2}) or (\ref{juold3}) (including the very specific case of a pure dark $Z$), with $(Q_Z)_V=(\frac{1}{2}-\,\sin^2\!\theta)\,Q-\frac{1}{4}\,(B\!-\!L)$ \linebreak in (\ref{qzu2},\ref{qzu8}) approaching the protophobic combination  \linebreak $\,[\,Q-\,(B\!-\!L)\,]/4\,$ for $\sin^2\theta$ close to 1/4.

\vspace{2mm}
We note that the $U$ charge 
 in (\ref{nphob}), corresponding to 
 \be
 {\cal J}^\mu_U\,=\,\frac{g"\cos\xi}{2}\ \left(\,J^\mu_A+J^\mu_{3L}-\sin^2\theta\,J^\mu_{\rm em}\right)\,
 \ee
 is,
for $\sin^2\theta$ close to 1/4, 
 \be
 \label{nphob1}
 Q_U\simeq\, \frac{1}{2}\, (F_A+T_{3L}-Q/4)\ \ \hbox{\it neutrino- and protophobic,}
 \ee
 in the sense indicated above.
Still our purpose is not to emphasize a $\,U\!$ uncoupled or little coupled to neutrinos or protons to avoid experimental constraints;
nor to overtry interpreting possible excesses as signs of a new boson. But rather to identify regions, \vspace{-.2mm}
in a multidimensional parameter space ($g",m_U,\,\gamma_Y,\gamma_B,\gamma_{L_i},\,\gamma_A,\,\eta, \,...)$,
for which various experiments may be sensitive, or not, to a $\,U$ boson, depending on its properties.

\vspace{-1mm}

\subsection{\boldmath A possible $U$ boson near 17 M\lowercase{e}V\,?$\!$?}

\vspace{-1mm}

The example of a $\approx 10$ MeV boson mediating the annihilations of light dark matter particles $\approx 4$ MeV was given in \cite{ldm1}, as an illustration of how the Hut-Lee-Weinberg lower mass bound of a few GeV's \cite{hut,lw} could be circumvented thanks to a new interaction, effectively stronger than weak interactions but only at lower energies.
Long before 2015, the Atomki group has been looking in $^8$Be transitions  for such a light $U$ boson
decaying into $e^+e^-$ \cite{kr}. Possible signs that might be interpreted as a neutral isoscalar axial-vector $U$ boson with $J^\pi\!=1^+$ already showed up in 2012 at $\approx 3\sigma$ \cite{kr2};  then at more than $5\sigma$ in 2015,  with a mass of 16.7~MeV$/c^2$ \cite{kr3}. Complementary indications appeared recently in $^4$He transitions \cite{kr4}.

\vspace{2mm}

A light $U$ with vector couplings \cite{U2,epjc}, taken as protophobic to satisfy $\pi^0 \!\to \gamma \,U$ decay constraints   \cite{NA48,NA482},
has been advocated as a possible interpretation of these events \cite{feng, feng2}, also subject to
the necessity of simultaneously avoiding too large effects in neutrino scatterings, and other requirements and constraints 
\cite{zm}, as from NA64 \cite{NA64}.
Other interpretations involving axial couplings have been discussed, as e.g.~in \cite{rose}.
The possibility of a $U$ boson with both axial and vector couplings, a naturally protophobic behavior originating  from the mixing with the $Z$  \cite{plb86},
and even reduced couplings to neutrinos as in (\ref{nphob0},\ref{nphob},\ref{nphob1}), may also be noted. 
Still more experimental information is desirable  before going too far in discussing interpretations for these results, which have not been confirmed independently yet.

\vspace{-1mm}

\subsection{Spin-dependent forces,\vspace{1.2mm} EP violations, \hbox{\ \ \ \ and a possible link with inflation}}
\vspace{-1mm}

Other possible manifestations of a spin-1 $U$ boson include a $CP$-conserving dipole-dipole interaction \cite{plb86} of the same type as for an axion \cite{mw}, thanks to eq.\,(\ref{gpga}), and 
a $CP$-violating mono\-pole-dipole interaction leading to a mass-spin coupling \cite{cqg,masspin}, both originating from an axial part in the $U$ couplings.

\vspace{2mm}
A very to extremely light (or even massless) $U$ boson with extremely small couplings to a combination of $B,\,L$ and $Q$ could lead to a new long-range force. By adding its effects to those of gravitation, it could be detected through apparent violations of the Equivalence Principle \cite{U2,plb86,adel,tou}. This constrains its gauge coupling to be extremely small, $\,g" \simle 10^{-24}$ \cite{micro}.

\vspace{2mm}
The extreme smallness of this gauge coupling may be related, within supersymmetric theories, with a very large extra-$U(1)$ abelian $\,\xi" \!D"$ term \cite{plb76,fi}, where the scale parameter $\sqrt {\xi"}$ is proportional to $1/\sqrt {g"}$. It corresponds to a huge vacuum energy density that may be at the origin of the very fast inflation of the early Universe, in connection with an extremely weak new long-range force, as already pointed out in \cite{sguts}.
The strength of the new force may then be related  with the inflation scale $\Lambda$ 
through a relation like
\be
g"\,\approx \,\left(\frac{m_{\rm sparticle}}{\Lambda}\right)^2,
\ee
leading to $g"$ of the order of  $\,\sim 10^{-24}$ for sparticle masses in the $\sim $ 1 to 10 TeV range with $\Lambda\sim 10^{15}$ to $10^{16}$ GeV \cite{micro}.

\vspace{2mm}

We shall now discuss in Sec.\,\ref{sec:u1} the general expression of the $F$ symmetry generator which may be gauged, in connection with the Brout-Englert-Higgs sector responsible for gauge symmetry breaking.
In Sec.\,\ref{sec:axial} we return to the simple case of $F$ as an axial generator $F_A$ as initially motivated by the 2-BEH structure of supersymmetric theories,  recalling how a longitudinal $U$ boson behaves much as an axionlike particle. This led us to make it very weakly interacting as in (\ref{rtheta}) thanks to a ``dark Higgs'' singlet $\sigma$, with the extra $U(1)$ broken at a higher scale.
We discuss in Sec.\,\ref{sec:general}
the mass of the $U$ boson and its mixing with the $Z$ occurring for $v_2\neq v_1$, leading to the $U$ current (\ref{juold2bis}).
We then extend these results
to an extra-$U(1)$ generated by an arbitrary  linear combination of the axial generator $F_A$ with $Y,\,B$ and $L_i$, plus a dark matter contribution, and a possible $\gamma_{F'}F'$ term acting on semi-inert BEH doublets uncoupled to quarks and leptons. We finally deal with grand-unification in Sec.\,\ref{sec:gut}.

\section{\boldmath Which \,extra-$\,U(1)$ \,symmetry?}
\label{sec:u1}
\vspace{-1.5mm}

\subsection{\boldmath A combination of $\,F_A,\,Y,\,B,\,L_i, \,F'$ \hbox{and $F_{\rm dark}$}}
\vspace{-2mm}

Which $U(1)$ symmetry generators may be gauged is restricted by the invariance of the Yukawa couplings responsible for quark and lepton masses, and depends on the number and characteristics of Brout-Englert-Higgs doublets, coupled or uncoupled to quarks and leptons.
With a single doublet, responsible for their masses as in the SM, any extra-$U(1)$ generator $F$ must act on quarks and leptons as a combination of the weak hypercharge $Y$ with baryon and lepton numbers $B$ and $L_i$, 
\be
\label{F0}
F\,=\, \gamma_B B+\gamma_{L_i} L_i + \gamma_Y Y+\gamma_d \,F_d\,.
\ee
The $U$ current is then obtained in the small mass limit, after mixing with the $Z$ current reconstructing the electromagnetic current as in (\ref{juold}-\ref{juold1bis}),  as a linear combination (\ref{juold1bis}) of the $B, \,L_i$ and electromagnetic currents with a possible dark matter current \cite{U2,epjc}.

\vspace{2mm}

There are additional possibilities with two or more BEH doublets.  We often concentrate on the remarkable situation for which two doublets, taken as 
\be
h_1=\left(\ba{c} h_1^0\, \vspace{1mm}\\ h_1^- \!\ea \right)\,,\ \ \ 
h_2=\left(\ba{c} h_2^+\!\vspace{1mm}\\ h_2^0\,  \ea \right)\,,\ 
\ee
give masses to  charged leptons and bottom quarks, and up quarks, respectively, through the usual trilinear Yukawa couplings as in supersymmetric extensions of the standard model, allowing for these theories to be made supersymmetric if desired \cite{plb76}.
These couplings, proportional to  $\,h_1\, \overline{ e_R\!\!}\ \,l_L, \ h_1\, \overline{ d_R\!\!}\ \,q_L $ and $h_2\, \overline{ u_R\!\!}\ \,q_L$ (+$\!$ h.c.), originate within supersymmetry from the trilinear superpotential
\be
\label{supot}
{\cal W}\,=\,\lambda_e \,H_1\ \overline{\!E}\,L \,+\,\lambda_e\,H_1\ \overline{\!D} \,Q \,+\,\lambda_u\, H_2 \ \overline{\!U}\,Q\,.
\ee
 Their gauge invariance requires
\be
\label{y}
\left\{\, \ba{ccc}
F(h_1) + F(l_L)- F(e_R) \!&= &\!0\,,
\vspace{2mm}\\
F(h_1) + F(q_L)- F(d_R) \!&= &\!0\,,
\vspace{2mm}\\
F(h_2) + F(q_L)- F(u_R) \!&= &\!0\,.
\ea \right.
\ee
This allows for the gauging of an axial $U(1)_A$ symmetry, with
\be
\label{fa}
\left\{\ba{ccc}
F_A(h_1)=F_A(h_2)=1\,,
\vspace{2mm}\\
F_A(q_L,\,l_L)= -\frac{1}{2}\ ,\ \ F_A(q_R,\,e_R)=\frac{1}{2}\,,
\ea \right.
\ee 
as done in the USSM with the $\mu\,H_2 H_1$ mass term replaced by a $\lambda\,H_2 H_1 \,S$ trilinear coupling with a singlet superfield $S$ having $F_A=-2$~\cite{plb76,npb75}. 
\vspace{2mm}

This gauging  of an extra-$U(1)$
may be done as well for $F(h_1)\!\neq F(h_2)$, i.e. with $F$ extended to a linear combination of $F_A$ and $Y$, possibly combined with $B$ and $L_i$. \linebreak
In fact, the seven values of $F$ for $(h_1,h_2;\,q_L,l_L;\,u_R,d_R,$ $e_R)$, for a single generation, related by the three eqs. (\ref{y}), depend on four parameters and may be expressed as 
$
F=\gamma_A F_A + \gamma_B B+\gamma_{L} L + \gamma_Y Y\,$.
Eqs.\,(\ref{y}), extended to the three families and taking into account CKM mixing angles between quarks, imply that the gauged quantum numbers of chiral quark and lepton fields, and of the two doublets $h_1$ and $h_2$, may be expressed as 
\be
\label{fabl}
F= \gamma_A F_A + \gamma_B B+\gamma_{L_i} L_i + \gamma_Y Y\,,
\ee
with 
\be
\label{fh}
\gamma_A\,=\,\frac{F(h_1) + F(h_2)}{2}\ , \ \   \gamma_Y\,=\, \frac{F(h_2) -F(h_1)}{2}\ .
\ee
This includes in particular couplings to
\be
T_{3R}=\,\frac{Y}{2} -\,\frac{B-L}{2}\,, \ \ \  \hbox{and/or} \ \ \ \frac{3B+L}{2}+ F_A\,,
\ee
which act only on right-handed quark and lepton fields. Even in these cases the resulting $U$ current, obtained as a combination of the extra-$U(1)_F$  and $Z$ currents, in general involves left-handed quarks and leptons as well through the term proportional to 
$(\gamma_Y+\eta)\,J^\mu_Z$ in eq.\,(\ref{curr40}), unless this one is absent as in (\ref{nomix},\ref{nomix2}).

\vspace{2mm}

We can still add extra terms 
to expression (\ref{fabl}) of $F$, provided they vanish for quarks, leptons and the BEH doublets responsible for their masses. This leads to
\be
\label{F}
\!F\,=\, \gamma_A F_A + \gamma_B B+\gamma_{L_i} L_i + \gamma_Y Y+\gamma_{F'} F'+ \gamma_d \,F_d\,.
\ee
$F_d$ acts only on particles and fields in a dark or hidden sector.
$F'$ may act on semi-inert BEH doublets $h'$, uncoupled to SM quarks and leptons but contributing to the mixing with the $Z$ as in eqs.\,(\ref{xi0},\ref{delta0}-\ref{etabis}).  

\vspace{2mm}

We may even consider identifying $F_d=\gamma_R R$ with the generator of continuous $R$-symmetry transformations  -- the progenitor of $R$-parity $R_p=(-1)^R$ in the supersymmetric standard model. Indeed $F$, including its possible $R$ contribution, is generally intended to be spontaneously broken so that the $U$ acquires a mass, then also allowing for the $R$-carrying gluinos and gravitino to acquire $U(1)_R$ breaking masses, after spontaneous supersymmetry breaking. $R$ does not act on SM particles but on their superpartners, including neutralinos and gravitinos which are potential dark matter candidates, then coupled to the $U$ boson which may act as the mediator of their annihilations.

\subsection{\boldmath Further possibilities for an extra $U(1)_F$}
\label{subsec:further}

\vspace{-1mm}

One can generalize further expression (\ref{F}) of $F$, still using conventional dimen\-sion-4 trilinear Yukawa couplings, by considering that up quarks, down quarks, and the three charged leptons may acquire their masses from up to five different \hbox{spin-0} BEH doublets (still allowing to avoid unwanted flavor-changing neutral current effects).
This would allow one to replace the single axial generator $F_A$ acting universally on quarks and leptons as in (\ref{fa}) by up to four different ones $F_{Aq}$ and $F_{Al_i}$ acting separately on quarks,  and on leptons of the three families,  leading to
\be
\label{F2}
\!F=\gamma_A F_{Aq} + \gamma_{Al_i}F_{Al_i}+ \gamma_B B+\gamma_{L_i} L_i + \gamma_Y Y+\gamma_{F'} F'+\gamma_d  F_d. 
\ee
\vspace{-3mm}

\noindent
 The resulting axial couplings are modified accordingly, those of down quarks and charged leptons
being no longer systematically equal as in (\ref{gaeb20},\ref{ga00},\ref{gaa0}).

\vspace{2mm}
One might go even further by generating quark and lepton masses through unconventional dimension $\geq 5 $ multilinear Yukawa couplings involving additional singlets charged under $U(1)_F$, so that eqs.\,(\ref{y}) would no longer have to be satisfied.  The $F$ values for $(h_1,h_2$; $q_L,l_L; \,u_R,d_R,e_R)$, and, subsequently, the resulting values for the $g_A$'s and $g_V$'s of quarks and leptons would then become unrelated parameters. There is also the possibility of flavor-changing neutral current couplings of the $U$ boson, however a potential source of difficulties.
Although we shall restrict  this study to the $F$ generator in (\ref{F}) for simplicity, it can be extended to more general situations. This is usually done, however, at the price of motivation, elegance, and predictivity, with {\it full generality implying a total loss of predictivity}.

\vspace{-1mm}

\subsection{\boldmath An extra singlet contributing to $m_U$}
\vspace{-1mm}

We also introduce a neutral singlet $\sigma$, intended to provide a small mass for the $U$ boson if the electroweak breaking is induced by a single BEH doublet (or two or more but with the same gauge quantum numbers) \cite{U2,epjc}; or to provide an extra contribution to its mass as in (\ref{mumu}) if the axial generator $F_A$ contributes to $F$, to make the longitudinal degree of freedom of a light $U$ with axial couplings sufficiently weakly interacting. This gauging involving $F_A$ originates from the supersymmetric standard model with a singlet superfield $S$ coupled  through a $\lambda\,H_2H_1S $ superpotential, with $F_A=1$ for $H_1$ and $H_2$ and $F_A=-\,2$ for the singlet $S$  \cite{plb76,npb75}.
One may also consider $S$ without coupling it to $H_1$ and $H_2$.
Independently of supersymmetry, one may consider a dimension-3 coupling  $\propto h_2h_1\,\sigma $ with $F_{A\sigma}=-\,2$, or a quartic coupling  $\propto h_2h_1\,\sigma^2$ with $F_{A\sigma}=-\,1$, or more generally consider $F_\sigma$ as an arbitrary parameter. 

\vspace{2mm}

We express the doublet and singlet  v.e.v.'s as
\be
\label{vv}
\left\{\
\ba{ccc}
\dis <h_1^0>\  \,=\,{v_1}/{\sqrt 2}\ ,\ \ \ <h_2^0> \ \,=\,{v_2}/{\sqrt 2}\ ,
\vspace{1mm}\\
\dis <\sigma>\  \,=\,{w}/{\sqrt 2}\ ,
\ea\right.
\ee
with
\vspace{-2mm}
\be
\label{vvb}
v_1=v\,\cos\beta\,,\ \ v_2=v\,\sin\beta\,,
\ee
where $\tan\beta =1/x=v_2/v_1$, and $v=2^{-1/4}\, G_F^{-1/2}\simeq 246$ GeV.
The $\,SU(2)\times U(1)_Y\,\times $ extra-$U(1)_F$ covariant derivative is given by
\be
\label{dcov}
iD_\mu = i\partial_\mu- (\,g\  \hbox{\boldmath $T$.\boldmath $W\!$}_\mu+\,\frac{g'}{2}\, Y\,B_\mu+\frac{g"}{2}\, F\,C_\mu\,)\,.
\ee
No ``kinetic-mixing'' term proportional to $B_{\mu\nu}\,C^{\mu\nu}$ needs to be considered, as it may be eliminated by returning to an orthogonal basis, with $C^\mu$ redefined as orthogonal to $B^\mu$. The $C^\mu$ coupling involves in general a contribution from the weak hypercharge current $J^\mu_Y$, already taken into account in expressions (\ref{F0}) or (\ref{F}) of $F$.

\vspace{-1mm}

\subsection{About anomalies}
\vspace{-1.5mm}

There are additional constraints from the requirement of anomaly cancellation. They allow for the gauging of $Y$, $B-L$ and $L_i-L_j$, in the presence of right-handed neutrino fields $\nu_R$. We may also consider axial generators
such as $F_A$ in (\ref{fa},\ref{F}), assuming anomalies to be cancelled, e.g. by mirror fermions, or exotic fermions similar to those in\,\,\,\underline{\hbox{\bf 27}}\,\,representations of $E(6)$ (decomposing into  \underline{\hbox{\bf 16}}+\underline{\hbox{\bf 10}}+\underline{\hbox{\bf 1}} =
\underline{\hbox{\bf 10}}+\underline{\hbox{$\bar{\bf 5 }$}}+\underline{\hbox{\bf 1}}\,+\,\underline{\hbox{\bf 5}}+\underline{\hbox{$\bar{\bf 5 }$}}\,+\,\underline{\hbox{\bf 1}}), or using other mechanisms.
In particular one may consider the same neutral currents as in $E(6)$, associated with $T_{3L},\,Y$, $[\,Y-\frac{5}{2}\, (B-L)\,]$ and $F_A$, and those associated with $L_i-L_j$. This leads to
\be
\label{F3}
\ba{c}
F\,=\,\gamma_A F_A + \gamma_{B-L} (B-L)+ \gamma_Y Y+\gamma_{F'}F' +\gamma_d \,F_d\vspace{2mm}\\
\hspace{8mm}+\,\delta_e\, (L_e- L_\tau) + \,\delta_\mu\,  (L_\mu-L_\tau)\,.
\vspace{-4.5mm}\\
\ea
\ee

\vspace{3mm}

\noindent
It corresponds to the earlier expression (\ref{F}) of $F$ with the relation $\sum_i \gamma_{L_i}=-\,3\,\gamma_B$.
Still we intend to perform the present study in a way as general as possible, irrespectively of the constraints from anomaly cancellation, which may involve presently unknown fermion fields. Or, otherwise, one may also choose to exclude the axial generator $F_A$ from the gauging by taking $\gamma_A=0$.

\section{\boldmath Starting with an axial $U$}
\label{sec:axial}

\vspace{-1.5mm}
\subsection{\boldmath \vspace{1mm}  $\nu$ scatterings and beam dump experiments}

\vspace{-2mm}

We now return for a short time to the simple gauging of 
\vspace{-.1mm} the axial generator $F=F_A$ in (\ref{fa}), choosing  $v_1=v_2=$ $v/\sqrt 2\,$ so that 
the extra $U(1)$ gauge field $C^\mu$ does not mix with the standard model one $Z^\mu_\circ=$ $\cos\theta \,W^\mu_3-\sin\theta \,B^\mu$. This  results in an axially-coupled $U^\mu\equiv C^\mu$, with ${\cal J}^\mu_U = (g"/2)\,J^\mu_A$ as from (\ref{juold}). The gauging of this extra $U(1)$, allowing to rotate independently $h_1$ and $h_2$ in supersymmetric theories, was originally done in connection with spontaneous supersymmetry breaking, triggered by the extra-$U(1)$ 
$\xi" D"$ term \cite{plb76}. This already provided a possible motivation for a very small coupling $g"$ and  a large supersymmetry-breaking scale from a very large $\xi"$ parameter inversely proportional to $g"$, associated with a huge vacuum energy density $\propto 1/g"^2$.

\vspace{2mm}
The left-handed and right-handed projectors  $(1-\gamma_5)/2$ and $(1+\gamma_5)/2$ are combined 
with  $F_A=-1/2$ and $+1/2$ for the chiral quark and lepton fields, respectively, as in (\ref{fa}).
The $U$ current  (including $g"/2$ from the covariant derivative in (\ref{dcov})) is expressed for quarks and leptons 
as
\be
\label{ju0}
{\cal J}^\mu_U\,=\,\frac{g"}{2}\,J^\mu_A\,=\,\frac{g"}{4}\ \hbox{\Large (} \!-\overline{\nu_L\!}\ \gamma^\mu\,\nu_L + \sum_{ude} \,  \bar f\,\gamma^\mu\gamma_5\,f  \, \hbox{\Large )} \,.
\ee
Without extra singlet v.e.v.~the $U$ mass is, for $v_1=v_2$,
\be
\label{mu}
m_U= \,\frac{g"v}{2}\,=\,\frac{g"}{g}\,m_W\,=\,\frac{g"}{\sqrt{g^2+g'^2}}\ m_Z\,,
\ee

\vspace{-1mm} 

\noindent
so that

\vspace{-6mm}

\be
\label{wzu}
\frac{G_F}{\sqrt 2}= \frac{g^2}{8m_W^2}=\frac{g^2+g'^2}{8m_Z^2}=\frac{g"^2}{8m_U^2}\ .
\ee
The neutral current effects of the $U$, for example in neutrino scattering experiments, are then generally too large if the $U$ is heavy.
If it  is light, however,
neutral current amplitudes, proportional to
\be
\label{prop}
\frac{g"^2}{8\, (m_U^2-q^2)}=\frac{G_F}{\sqrt 2}\ \frac{m_U^2}{m_U^2-q^2}\simeq -\,\frac{g"^2}{8q^2}\ \ \,\hbox{for}\ 
\ m_U^2\ll |q^2|\,,
\ee
can be sufficiently small if $g"$ is small enough, with the $U$ light as compared to $|q^2|$ in the experiment considered,
providing an early motivation for making a new neutral gauge boson light (cf. subsection \ref{subsec:tau}).

\vspace{2mm}
The gauge coupling $g"$ and axial couplings to quarks and charged leptons may  then be expressed proportionally to $m_U$, reading in this first simple situation as
\be
\ba{ccl}
\label{ga-1}
\dis g_A\,=\,\frac{g"}{4}\!&=&\! \dis \frac{g}{4}\ \frac{m_U}{m_W}
\,=\, \frac{m_U}{2v}\,=\,2^{1/4}\,G_F^{1/2}\ \frac{m_U}{2}
\vspace{3mm}\\
&\simeq&\ \ \ \ 2\times 10^{-6}\ m_U\rm (MeV)\,.
\ea
\ee
This benchmark value appears in many discussions of a new light gauge boson as a mediator of non-standard neutrino interactions; or produced in meson decays or beam dump experiments, or contributing to atomic-physics parity violation, etc..
Its axial couplings are then expressed as in (\ref{gaa0}) as a fraction of this benchmark value, according to
\vspace{-2mm}
\be
\label{gaa-1}
 g_{A\,\pm}\,\simeq \,2\times 10^{-6}\ m_U{\rm (MeV)}\ A_\pm\,.
 \ee
 
\vspace{1mm}

Such a $U$ boson axially coupled with full strength as in (\ref{ga-1}), i.e. with $A_\pm =1$,
was constrained long ago to be heavier than 7 MeV to be sufficiently short-lived in beam dump experiments
\cite{U}, with a decay rate $\,\Gamma\propto g"^2 \,m_U\propto m_U^3$, and a
decay length   
$l =\beta\gamma \, c\tau\,\propto\, g"^{-2}\,m_U^{-2}\propto m_U^{-4}$ (cf. subsection \ref{subsec:tau}).
It had also to be  lighter than 300 MeV, to sufficiently benefit from propagator effects in neutrino-electron scattering experiments for $m_U^2$ smaller than the relevant $|q^2|$ under consideration at the time.
This left, forty years ago, a window of opportunity between 7 and 300 MeV for an axially coupled boson, in the absence of an extra singlet allowing for $|A_\pm|$ to be significantly smaller than 1.
This early example illustrates the complementary roles of neutrino scatterings at lower $|q^2|$, and beam dump experiments, in the search for new light gauge bosons.
\vspace{-2mm}

\subsection{\boldmath The need for an invisible axionlike behavior}
\label{subsec:axionlike}

\vspace{-2mm}

The longitudinal degree of freedom of an ultrarelativistic  $U$ boson with axial couplings 
would behave much as an axionlike particle, with effective pseudoscalar couplings to quarks and leptons given by eq.\,(\ref{gpga}). With $g_A=g"/4$ as in (\ref{ga-1}) they would read, for equal doublet v.e.v.'s,
\be
\dis g_{P\,ql}\,=\, \frac{2m_{ql}}{m_U}\,g_A =2^{1/4}\,G_F^{1/2}\ m_{ql}\simeq 4\times 10^{-6}\, m_{ql}\, \rm (MeV) \ .
\ee
reconstructing the pseudoscalar couplings of a standard axion $A$, for $v_2=v_1$ \cite{w1,w2}.

\vspace{2mm}

We thus considered very early a large singlet v.e.v. $<\sigma>\ = w/\sqrt 2\,$,
increasing $m_U$ to
\be
\label{mu2}
m_U\,=\, \frac{g"\!}{2}\  \sqrt{v^2+F_\sigma^2 w^2}\,=\,\frac{g"}{g}\ \frac{m_W}{r}\,,
\ee
defining the invisibility parameter \cite{U}
\be
\label{rA}
r=\cos\theta_A= \frac{v}{\sqrt{v^2+ F_\sigma^2 w^2}}\ \leq 1 \ \,,
\ee
the equivalent pseudoscalar  $a$  in (\ref{rtheta}) being mostly an electroweak singlet for small $r\!=\!\cos\theta_A$, with
$\tan\theta_A = |F_\sigma| \,w /v$.

\vspace{2.5mm}

The axial coupling $g_A=g"/4$ in (\ref{ju0}) gets expressed proportionately to $m_U$ and $r$, 
with
\be
\label{gar}
g_A\,=\,\dis\frac{g"}{4}=\,\dis \frac{g}{4}\  \frac{m_U}{m_W}\ r
\simeq \,2\times 10^{-6}\ r \ m_U\rm (MeV)\,.
\ee
The effective pseudoscalar couplings (\ref{gpga}) of a longitudinal $U$,
\vspace{-1.5mm}
\be
\label{formr}
\ba{ccl}
\dis g_{Pql} \,=\,  g_A\, \frac{2\,m_{ql}}{m_U}\!&=&\ \  \dis 2^{1/4}\ G_F^{1/2}\ r\ m_{ql}
\vspace{1mm}\\
\!&\simeq &\dis 4\times 10^{-6}\  r\ m_{lq}\rm (MeV)\, ,
\ea
\ee
get proportional to $r=\cos\theta_A$, in agreement with (\ref{rtheta}).
For a given $m_U$, the amplitudes involving a $U$, proportional to $g_V,\,g_A$ or $g_P$ and thus to $r$, are sufficiently small  if $r$ is small enough, i.e.~$|F_\sigma | w $ large enough compared to the electroweak scale.
The $U$ decay rate gets reduced, leading to a lifetime and decay length increased by a factor $1/r^2$,
\be
\left\{\ba{ccccc}
\tau\!&\propto&\! \,g"^{-2}\,m_U^{-1}\!&\propto&\! m_U^{-3}\,r^{-2}\,,
\vspace{2mm}\\
l =\beta\gamma \, c\tau\!&\propto&\!  g"^{-2}\,m_U^{-2}\!&\propto&\!  m_U^{-4}\,r^{-2}\,,
\ea\right.
\ee
\vspace{-1mm}

\noindent
allowing for a sufficiently weakly coupled $U$ boson of any mass to satisfy direct experimental constraints, including from beam dump experiments.

\vspace{-1.5mm}

\subsection{``Invisibility''  \vspace{1.5mm}and high scale symmetry \hbox{or supersymmetry} breaking}

\vspace{-1.5mm}

This mechanism relying on a large singlet v.e.v. $w/\sqrt 2$ making the $U$ boson almost ``invisible'' also led us independently \cite{U} to the ``invisible axion'' mechanism \cite{dfs,z} for an anomalous global $U(1)$.

\vspace{2mm}

We had already used this property of a very light gauge boson behaving as a Goldstone boson (as in the standard model in the limit $g,g'\to 0, m_W,m_Z\to 0$ with $v$ fixed), and interacting with amplitudes inversely proportional to the symmetry breaking scale. It provided a laboratory to establish this equivalence property for supersymmetric as well as for gauge theories, with, furthermore, the gauge particle, either fermionic or bosonic, interacting quasi-``invisibly'' if the supersymmetry-breaking or gauge-symmetry breaking scale is large  \cite{grav}.

\vspace{2mm}

A very light \hbox{spin-3/2} gravitino 
then behaves very much as the corresponding (``eaten-away'') spin-1/2 goldstino.  It can also be made almost ``invisible'', i.e. with very reduced interactions $\propto 1/d$ or $1/F$, thanks to a very large supersym\-metry-brea\-king scale parameter ($d$ or $F$), i.e. a very large auxiliary-field v.e.v. within an extra-$U(1)$, dark or hidden sector. This also provides an early decoupling of a very-weakly interacting gravitino.
For an extra-$U(1)$  this is associated with a very small gauge coupling $g"$ (and effective couplings of the equivalent goldstino) \cite{plb76,sguts,micro}.
Supersymmetry, spontaneously broken ``at a high scale'' in a hidden sector, gets softly broken in the visible one.

\section{\boldmath $U$ mass and mixing angle with $\,Z$}
\label{sec:general}

\vspace{-1mm}

\subsection{\boldmath $U$ as an axial boson mixed with the $Z$}

\vspace{-1.5mm}

We now discuss more precisely mixing effects between the extra-$U(1)$  gauge field $C^\mu$ and the $Z_\circ^\mu$ of the standard model, when the two doublets have different v.e.v.'s so that the $U$ current picks up an extra term proportional to $J^\mu_Z=J^\mu_{3L}-\sin^2\theta \,J_{\rm em}^\mu$, proportionally to $(v_2^2-v_1^2)/v^2=-\,\cos 2\beta$, as expressed in eqs.\,(\ref{juold},\ref{juold2bis},\ref{juold5}).
\,We first consider the gauging of the axial generator $F_A$, with $h_1$ and $h_2$ having $Y=\mp\, 1$ and  $F\!=\!F_A\!=\!1$, and a singlet $\sigma$ (now often referred to as a ``dark Higgs'' field). 
When  $h_1$ and $h_2$ acquire different v.e.v.'s $C^\mu$ mixes with the $Z^\mu_\circ$  of the standard model
through the mass$^2$ terms 
\be
\label{lm}
\ba{ccc}
{\cal L}_m \!&=&\dis \frac{(g_Z Z_\circ ^\mu+g" C^\mu)^2 \, v_1^2}{8}\, +\, \frac{(-\,g_Z Z_\circ ^\mu+g" C^\mu)^2 \, v_2^2}{8} 
\vspace{2mm}\\
&& + \dis \frac{g"^2F_\sigma^2w^2}{8}\ C_\mu C^\mu\,,
\vspace{-6mm}\\
\ea
\ee
\vspace{0mm}

\noindent
with $g_Z^2=g^2+g'^2$. This includes a $C^\mu$ mass term
$
 m_C = g" |F_\sigma| w/{2}$,
induced by $<\!\sigma\!>\ = w/\sqrt 2$,
and leads to the mass$^2$ matrix \cite{plb86}
\bea
\label{m2}
\ba{ccc}
{\cal M}^2= \dis \frac{1}{4}\left(\ba{ccc}  
g_Z^2\,(v_1^2+v_2 ^2)&&  -\,g_Z g"\,(v_2^2-v_1^2)
\vspace{2mm}\\
-\,g_Z g"\,(v_2^2-v_1^2)  && g"^2 \,(v_1^2+v_2 ^2+F_\sigma^2 w^2)
\ea\right)
\ea
\nonumber
\eea
\be
\ba{ccc}
\!=\dis \frac{v^2}{4} \left(
\ba{ccc}
 g_Z^2&&  g_Z g"\cos 2\beta
\vspace{2mm}\\
g_Z g"\cos 2\beta  && \dis g"^2 \,(\cos^2 2\beta\!+\!\sin^2 2\beta +\frac{F_\sigma^2 w^2}{v^2})
\ea\!\right)\!.
\vspace{1mm}\\
\ea
\ee

\vspace{2mm}

Its mass eigenstates are 
\be
\label{mix}
\left\{\ba{ccc}
Z^\mu \!&=&\! \cos\xi \ Z_\circ^\mu\, -\,\sin\xi \ C^\mu\,,
\vspace{2mm}\\
U^\mu\!&=& \sin\xi \ Z_\circ^\mu  \,+\cos\xi \ C^\mu\,,
\ea\right.
\ee
with 
\be
\label{mix2}
\left\{\ba{ccc}
Z_\circ^\mu \!&=&\! \cos \theta \ W_3^\mu-\sin\theta \ B^\mu\,,
\vspace{2mm}\\
A^\mu  \!&=&\! \sin \theta \ W_3^\mu+\cos\theta \ B^\mu\,,
\ea\right.
\ee
as in the standard model, so that
\be
\label{mix3}
\left(\ \ba{c}
Z^\mu
\vspace{1mm}\\
A^\mu
\vspace{1mm}\\
U^\mu
\ea\right) = 
\left(\ \ba{ccc}
c_\xi c_\theta & - c_\xi s_\theta & \ -s_\xi
\vspace{1mm}\\
\ \ \ s_\theta & \ \ \ \ c_\theta & \ \ 0
\vspace{1mm}\\
s_\xi c_\theta& - s_\xi s_\theta & \ \ \,c_\xi
\ea\ \right) 
 \left(\ \ba{c}
W_3^\mu
\vspace{1mm}\\
B^\mu
\vspace{1mm}\\
C^\mu
\ea\right) .
\ee
The $Z$-$U$ mixing angle $\xi$ is given by
\be
\label{mixex}
\tan 2\xi \, =\, \frac{2g_Zg" (v_2^2-v_1^2)}{(g_Z^2-g"^2)(v_1^2+v_2^2)-g"^2F_\sigma^2w^2}\ .
\ee
In the small $m_U$ limit and considering the contribution  from the last two terms in the $CC$ matrix element in (\ref{m2}) as a perturbation, $\xi$ is approximately given by \cite{plb86}
\be
\label{tanxi}
\tan\xi\,\simeq\, \frac{g"}{g_Z}\ \frac{v_2^2-v_1^2}{v_2^2+v_1^2} \,=\,   \frac{g"}{g_Z}\ (-\,\cos 2\beta)\,.
\ee
This is a special case of the general formula (\ref{xi0}) for any set of BEH doublets, referred to as $\varphi_i$ with $Y=+1$, or equivalently $h_i$, with $Y =\epsilon_i=\pm 1$   \cite{U2},
\be
\label{tanxi2}
\!\!\ba{ccl}
\tan\xi \!&\simeq& {\dis \!\frac{g"}{g_Z}}\, \sum_i F(\varphi_i)\  {\dis \frac{v_i^2}{v^2}\,=\, \frac{g"}{g_Z}}  \,\sum_i \, \epsilon_i \,F(h_i)\ \dis \frac{v_i^2}{v^2}\ .
\ea
\ee

\vspace{2mm}

Taking the last two terms in the $CC$ matrix element of ${\cal M}^2$  as a perturbation, we identify 
\be
\label{mzu0}
m_Z^2\,\simeq \, \frac{(g^2+g'^2)\,v^2}{4\,\cos^2 \xi}
\,=\, \frac{(g^2+g'^2+g"^2 \cos^2 2\beta)\,v^2}{4}\,,
\ee
then get, from these last two terms,
\be
\label{mu20}
\ba{ccl}
m_U^2
\!&\simeq&\! \dis \frac{g"^2\cos^2\xi}{4}  \left(v^2\sin^2 2\beta \!+\! F_\sigma^2 w^2\right)
\ea
\ee
or
\be
m_U\,=\,\dis  \ \frac{g"\cos\xi}{2} \ \frac{v\sin 2\beta}{r}\,.
\ee
We have defined, as in (\ref{rA}) for $\beta=\pi/4$,  the invisibility parameter
\be
\label{r2}
r=\cos\theta_A\,=\,\frac{v\,\sin 2\beta}{\sqrt{v^2\,\sin^2 2\beta + F_\sigma^2 w^2}}\ ,
\ee
with
$\,\tan\theta_A= (|F_\sigma | w)/(v\,\sin 2\beta)$\,. 
$g"$ can be expressed proportionally to  $m_U$ and $r$ as in (\ref{valga}), with
\be
\label{g"mur2}
\framebox [7cm]{\rule[-.35cm]{0cm}{1cm} $ \dis
\frac{g"\cos\xi}{4}\,\simeq\,2\times 10^{-6}\,m_U(\hbox{MeV})\ \frac{r}{\sin 2\beta}\ ,
$}
\ee
where $r$ is small when $|F_\sigma| w$ is large compared to $v\,\sin 2\beta$.
$g"/4\simeq 2\times 10^{-6}\,m_U(\hbox{MeV})$ occurs as the natural benchmark value for the extra-$U(1)$ gauge coupling, as seen for non-standard neutrino interactions in subsections \ref{subsec:mec} and \ref{subsec:mec2}.
Altogether we have in this case
  \be
  \label{sum}
 \left\{\ \ba{ccc}
  \dis \tan\xi\!&\simeq &\! \dis
-\  \frac{g"}{\sqrt{g^2+g'^2}}\ \cos 2\beta\,,\ \ \ \ \ \ \ \ \ \ \ \ 
 \vspace{3mm}\\
  m_Z\!&\simeq&\!\dis \frac{\sqrt{g^2+g'^2+g"^2\,\cos^2 2\beta}\ v}{2}\,\simeq\,\frac{m_{Z_\circ}}{\cos\xi}\ ,
    \vspace{2.5mm}\\
    m_U\!&\simeq &\, \dis \frac{g"\cos\xi }{2}\ \sqrt{\, \sin^2 2\beta \,v^2+  F_\sigma^2\,w^2}\ ,
      \vspace{2.5mm}\\
  r    \!  &=& \!\!\!\! \cos\theta_A \ \simeq \  \dis  \frac{v\, \sin 2\beta}{\sqrt{\,v^2\, \sin^2 2\beta+  F_\sigma^2\,w^2}}\ .
 \ea \right.
 \ee

\vspace{1mm}
 
Eqs.\,(\ref{valga},\ref{g"mur2}) determining the gauge coupling $g"$ in terms of $m_U$, the invisibility parameter $r=\cos\theta_A$ and the ratio of spin-0 v.e.v.'s  $\,\tan \beta$, \,in a two-doublet + one-singlet model, remain valid under more general circumstances, through essentially the same calculation.

\vspace{-1mm}

\subsection{\boldmath \vspace{1.5mm}General case\hspace{-.5mm}: \hbox{$U$ as an extra $U(1)_F$ gauge boson mixed with $Z$}}

\vspace{-.5mm}

The extra-$U(1)$ generator $F$ may involve as in (\ref{F}), in addition to the axial  $F_A$, terms proportional to $Y, \ B$ and $L_i$,  a (light) dark matter contribution $\gamma_d\,F_d$, and a $\gamma_{F'}F'$ term acting on semi-inert doublets uncoupled to quarks and leptons. 
There is no need to consider a  kinetic-mixing term $\propto B_{\mu\nu} C^{\mu\nu}$, immediately eliminated by redefining $C^\mu$ as orthogonal to $B^\mu$.

\vspace{2mm}
For $h_1$ and $h_2$ taken with $Y=\mp 1$ and  $F=F_i$, and a singlet $\sigma$ with $F=F_\sigma$.
the mixing angle $\xi$ is given in the small $g"$ limit by  (\ref{xi0}),
\be
\label{tanxif}
\tan\xi\,\simeq\,\frac{g"}{g_Z}\ (F_2\,\sin^2\beta - F_1\,\cos^2\beta)\,.
\ee
The quantum numbers of $\,h_1,h_2$ and $\sigma$ are 
\be
\label{fh2}
\left\{\ 
\ba{ccl}
 F_1\!&=&\gamma_A\, - \, \gamma_Y\ ,
\vspace{1.5mm}\\
F_2\!&=&\gamma_A \, +\, \gamma_Y\ ,
\ea \right.\ \ \ 
F_\sigma\,=\,\gamma_A\,F_{A\sigma}+\gamma_d\,F_{d\sigma}\ .
\ee
This allows  for $\sigma$ to transform under $U(1)_A$ and/or under a dark $U(1)_d$ symmetry generated by $F_d\,$.
(The calculation, formulated with $F=\gamma_A F_A+ \gamma_Y Y= \gamma_Y\mp \gamma_A$ for $h_1^c$ and $h_2$, applies as well for a SM-like doublet $h_{\rm sm}$ and a semi-inert one $h'$, with $\,F=\gamma_Y Y+\gamma_{F'}\,F'$ taken as $\gamma_Y$ and $\gamma_Y +2$, respectively.)

\vspace{2mm}
The mass terms (\ref{lm}) read
\be
\label{lm2}
\ba{ccl}
\!\!{\cal L}_m 
\!&=&
 \dis \frac{1}{2}\ [\,\,g_Z \,Z_\circ ^\mu- g"  (\gamma_Y - \gamma_A \cos 2\beta )\,C^\mu\,]^2 \  \frac{v^2}{4} 
\vspace{2mm}\\
&&\!\dis +\ \frac{1}{2}\ \,\frac{g"^2}{4} \ 
(\gamma_A^2\sin^2 2\beta \,v^2 + F_\sigma^2\,w^2) \ C_\mu C^\mu\,,
\vspace{.5mm}\\
\ea
\ee

\vspace{0mm}

\noindent
and the mixing angle $\xi$ is  given in the small mass limit by (\ref{tanxif}), which reads
\be
\label{tanxi3}
\tan\xi 
\,\simeq\,\frac{g"}{g_Z}\ (\gamma_Y-\,\gamma_A\cos 2\beta)\,.
\ee
The first mass$^2$ term in (\ref{lm2}) leads to 
\be
\label{mz4}
\ba{ccl}
m_Z^2\!&\simeq&\ \  \dis \,\frac{g^2+g'^2}{4\,\cos^2\xi}\ v^2
\vspace{2mm}\\
&\simeq&\,\dis \frac{g^2+g'^2+g"^2\,(\gamma_Y-\,\gamma_A\cos 2\beta)^2}{4}\ v^2\ .
\ea
\ee
The second one,  $\propto C_\mu C^\mu=\, (\cos\xi \ U^\mu-\,\sin\xi \ Z^\mu)^2 $ and considered as a perturbation, leads to
 \be
  \label{mu4}
 \dis  m_U^2\,\simeq\ \frac{g"^2\cos^2\xi}{4}\ ( \gamma_A^2\,\sin^2 2\beta\,v^2+  F_\sigma^2 w^2)\,.
  \ee
 It is independent of $\gamma_Y$, 
 and reduces to (\ref{mu20}) for $\gamma_A=1$. \linebreak
  We define as in (\ref{rtheta},\ref{r2})  $r=\cos\theta_A$, 
  given by
  \be
  \tan\theta_A\,=\,\frac{|F_\sigma|\,w}{|\gamma_A|\,\sin 2\beta\,v}\,.
  \ee
  
\vspace{1mm}

We summarize these results on the $U$ mass, mixing angle $\xi$ and invisibility parameter $r$ by

\vspace*{-2mm}
  \be
  \label{sum2}
  \framebox [8.6cm]{\rule[-3.2cm]{0cm}{6.6cm} $ \dis
\!\! \ba{ccc}
  \dis \tan\xi\!&\simeq &\! \dis
 \frac{g"}{\sqrt{g^2+g'^2}}\ \,(\gamma_Y-\,\gamma_A\cos 2\beta)\,,\ \ \ 
  \vspace{2mm}\\
   \dis \cos^2\xi \!\!&\simeq &\! \dis \frac{g^2+g'^2}{g^2+g'^2+g"^2\, (\gamma_Y-\,\gamma_A\cos 2\beta)^2}\ ,
  \vspace{3.5mm}\\
  m_Z\!&\simeq&\!\dis\! \frac{\sqrt{g^2\!+\!g'^2\!+g"^2\,(\gamma_Y\!-\gamma_A\cos 2\beta)^2}\ v}{2}\simeq\frac{m_{Z_\circ}}{\cos\xi}\, ,
    \vspace{2.5mm}\\
    m_U\!&\simeq &\! \dis\frac{g"\cos\xi}{2}\,\sqrt{\gamma_A^2\, \sin^2 2\beta \,v^2+  F_\sigma^2\,w^2}\,,
        \vspace{1.5mm}\\
     &\simeq&\!\dis\hspace{-16mm} \frac{g"\cos\xi}{2}\  |\gamma_A| \ \,\frac{v\,\sin 2\beta}{r}\,,
      \vspace{2.5mm}\\
     r\!   &=&\hspace{-8mm}\cos\theta_A \ = \  \dis  \frac{|\gamma_A|\,v\, \sin 2\beta}{\sqrt{\gamma_A^2\,v^2\, \sin^2 2\beta+  F_\sigma^2\,w^2}}\ .
 \ea
 $}
 \ee
 
 \vspace{4mm}
For $\gamma_A\!\neq0$ and normalizing $g"$ to $\gamma_A=1$, 
 this reduces for $\gamma_Y=0$ to  (\ref{sum}), the $U$ being coupled as in (\ref{ju0}) to an axial current, plus a contribution induced by the mixing with $Z$, for $\tan\beta\neq1$.
For $\beta=\pi/4$ there is no mixing and we recover $m_Z=g_Z v/2$ 
\vspace{-.2mm}
as in the SM and
$m_U = g"\sqrt{v^2+ F_\sigma^2 w^2}/2$ $= g" v/(2r)$ as in (\ref{mu2}).
In the other limit $\sin 2\beta \!\to\! 0$, e.g. through a small $v_1$ with $\beta\simeq \pi/2$ and $\cos 2\beta\simeq -1$, 
\vspace{-.2mm}
 we approach a one-doublet-v.e.v. situation with 
$m_Z\simeq \sqrt {g^2+g'^2+g"^2}\ v/2\,$, $\,m_U\simeq g"\cos\xi \,|F_\sigma|\, w/2$, and vector couplings of the $U$ to massive up quarks.
 \vspace{2mm}
 
For $\gamma_A\neq0$ and normalizing $g"$ to $\gamma_A=1$, we have again as in (\ref{g"mur2}), independently of $\gamma_Y,\,\gamma_B$ and $\gamma_{L_i}$,
\be
\label{g"mur3}
\frac{g"\cos\xi}{4}\,\simeq\,2\times 10^{-6}\,m_U(\hbox{MeV})\ \frac{r}{\sin 2\beta}\ ,
\ee
 with $\eta=-\,\cos2\beta$\,.

 \vspace{2mm}

When $F_A$ does not participate in the gauging so that $\gamma_A=0$,
 and normalizing $g"$ to $\gamma_Y\!=1$,  the $U$ would remain massless without the singlet $\sigma$. Then we have
    \be
  \label{sum3}
\left\{\  \ba{ccl}
  \dis \tan\xi &\simeq &\  \dis
 \frac{g"}{\sqrt{g^2+g'^2}}\ ,
    \vspace{2mm}\\
  m_Z\!&\simeq&\!\dis \frac{\sqrt{g^2+g'^2+g"^2}\ v}{2}\ ,
    \vspace{2mm}\\
    m_U\!&\simeq &\! \dis\frac{g"\cos\xi \ |F_\sigma| \ w}{2}\,,
 \ea \right.
 \ee
the 3 x 3 mixing matrix (\ref{mix3}) reducing to  \cite{U2,epjc}
 \be
\label{33}
\left\{\, \ba{ccl}
Z\!&\simeq&\!  \hbox{$\dis \frac{g \,W_3-g'B-g''C}{\sqrt{g^2+g'^2+g''^2}}$} \ ,
\vspace{2mm}\\
A\!&=&\!\ \  \hbox{$\dis \frac{g' \,W_3+g\,B}{\sqrt{g^2+g'^2}}$}\ ,
\vspace{2mm}\\
U\!&\simeq &\!  \hbox{$\dis \frac{g''(g \,W_3-g'B)+(g^2+g'^2)\,C}{\sqrt{g^2+g'^2}\,\sqrt{g^2+g'^2+g''^2}}$}\ ,
\ea
\right.
\ee

\noindent
with the $U$ vectorially coupled to quarks and leptons.

\subsection{\boldmath Exact determinations of $m_U$ and $\xi$}

\vspace{-1mm}

The mass$^2$ matrix corresponding to (\ref{lm2}), which includes  the element
\be
\ba{ccl}
\!m_{CC}^2\!\!&=&\! \dis \frac{g"^2}{4} \ [\,(\gamma_A-\gamma_Y)^2\,v_1^2+ (\gamma_A+\gamma_Y)^2\,v_2 ^2+F_\sigma^2 w^2\,] 
\vspace{2mm}\\
&& \hspace{-8mm}
= \ \dis \frac{g"^2v^2}{4} \,[\,(\gamma_Y-\gamma_A\cos 2\beta)^2+\gamma_A^2\sin^2 2\beta +\frac{F_\sigma^2 w^2}{v^2}\,]\,,
\vspace{2mm}\\
\ea
\ee

\noindent
reads
\be
\label{m2H}
{\cal M}^2=\frac{v^2}{4}
\left(\!\ba{cc}
g_Z^2 & \ \ \ 
-g_Zg"\,(\gamma_Y+\eta)
\vspace{3mm}\\
-g_Zg"\,(\gamma_Y+\eta) & \ \ \ g"^2\, [\,(\gamma_Y+\eta)^2+H\,]
\ea\!\right)\!,
\ee

\vspace{1mm}
\noindent
where $\,\gamma_Y+\eta = F_2\sin^2\!\beta - F_1\cos^2\!\beta=\gamma_Y-\gamma_A\cos2\beta$, with
\be
H\,= \,\gamma_A^2\,\sin^2 2\beta\, +\,\frac{F_\sigma^2 w^2}{v^2}\ .
\ee

\vspace{2mm}

$m_Z^2$ and $m_U^2$ are obtained from their product and sum, 
\be
\label{sp3}
\left\{\
\ba{ccl}
\Pi &=& \dis \frac{(g^2 +g'^2)\,v^2}{4}\ \frac{g"^2v^2}{4}\ H\ ,
\vspace{2mm}\\
\Sigma &=&\dis \frac{[\,g^2+g'^2+g"^2\,(\gamma_Y+\eta)^2+H)\,]\,v^2}{4}\ ,
\ea\right.
\ee

\noindent
so that
\be
\label{sp2}
m_Z^2\,,\,m_U^2 \,=\,
\frac{\Sigma}{2}\,\left( 1 \pm\sqrt{1-\frac{4\,\Pi}{\Sigma^2}}  \ \right)\,,
\ee
also valid for a heavy $U$. The mixing angle $\xi$ is given by
\be
\tan2\xi\,=\,\frac{2g_Zg"\,(\gamma_Y+\eta)}{g_Z^2-g"^2\,[(\gamma_Y+\eta)^2+H]}\,,
\ee
reducing to (\ref{mixex}) for $\gamma_Y=0,\  \gamma_A=1$ and $\eta=-\cos2\beta$,
 and to (\ref{tanxi3}) for small $m_U$, with $\eta=-\,\gamma_A\cos 2\beta$.

\vspace{2mm}

These formulas giving the exact values of $m_U$ and $\xi$ in a general two-doublet + one-singlet situation  may also be applied to not-so-light or even heavy $U$ bosons, that could be detected through their decays $U\to \mu^+\mu^-$, as in LHCb for example \cite{lhcb1,lhcb2}.

\vspace{2mm}

$m_Z^2$ and $m_U^2$  can be developed in the small $m_U$ limit, according to
\be
\label{sp2bis}
\left\{ \ba{ccc}
m_Z^2 \!&=& \dis \Sigma\,-\,\frac{\Pi}{\Sigma}\, -\,\frac{\Pi^2}{\Sigma^3} \,+\, ...\ ,
\vspace{2mm}\\
m_U^2  \!&=& \dis \frac{\Pi}{\Sigma} \,+\, \frac{\Pi^2}{\Sigma^3}\, +\,2\ \frac{\Pi^3}{\Sigma^5}\,+\,  ...\ .
\ea \right.\,
\ee
A first approximation,
\be
m_U^2\,\simeq\,\dis \frac{\Pi}{\Sigma}\,\simeq\, \frac{g"^2v^2}{4}\ H
\ee
leads to
\be
\label{mz0}
\ba{ccl}
\dis m_Z^2\,\simeq\, \Sigma-\frac{\Pi}{\Sigma}\!&\simeq&\! \dis\frac{[\,g^2+g'^2+g"^2\,(\gamma_Y+\eta)^2]\ v^2}{4}
\vspace{1mm}\\
&\simeq&\ \ \ \dis \frac{m_{Z_\circ}^2}{\cos^2\xi}\ ,
\vspace{-2mm}\\
\ea
\ee
and at next order to
\be
\ba{ccl}
m_U^2\!&\simeq&  \dis\dis\frac{\Pi}{\Sigma-\frac{\Pi}{\Sigma}}\,\simeq\,  \dis\cos^2\xi\ 
\frac{g"^2v^2}{4}\ H\,
\vspace{2mm}\\
\!&\simeq&  \dis\frac{g"^2\cos^2\xi}{4}\ ( \gamma_A^2\,\sin^2 2\beta\,v^2+  F_\sigma^2w^2)\,.
\ea
\ee

\vspace{2mm}

\noindent
It gives back $m_Z^2$ and $m_U^2$ in (\ref{mz4},\ref{mu4}), in which $m_U^2$ could be directly obtained  as 
\be
m_U^2\,\simeq\, \cos^2\xi\,\left(m_{CC}^2-\frac{(m^2_{Z_\circ C})^2}{m^2_{Z_\circ Z_\circ}}\right)\,\simeq\,\cos^2\xi\ \frac{g"^2v^2}{4}\ H\,.
\ee
There is no need to consider further corrections  if the $U$ is light, $\Pi^2/\Sigma^4\simeq m_U^4/m_Z^4$ being very small.

\vspace{3mm}

With
\vspace{-.1mm}
 $\gamma_A=(F_2+F_1)/2=
 [F(h_2)-F(h_1^c)]/2$\,, 
this expression of $m_U$ reads, 
in any two-doublet + singlet situation
where the two doublets (called $h_1^c$ and $h_2$, or $h$ and $h'$)  have the same $Y=1$,
\be
\label{mu2h}
\!m_U\,\simeq\,\frac{g"\cos\xi}{2}\ \sqrt{\, \left(\frac{F(h')-F(h)}{2}\right)^{\!2}\, v^2\sin^2 2\beta +F_\sigma^2 w^2}\,.
\ee
This also illustrates how two doublets with the same $F$ and $Y$  act as a single one, breaking only one neutral gauge symmetry generator, thus providing no contribution to $m_U^2$.

\vspace{-2mm}

\subsection{\boldmath \vspace{1.5mm}Application to a semi-inert doublet $h'$}
\label{subsec:inert}

\vspace{-2mm}

In particular, let us consider a SM-like doublet $h_{\rm sm}$ 
with v.e.v. $v\cos\beta'/\sqrt 2$ and $F\!=\!\gamma_Y$, 
\vspace{-.5mm}
and a semi-inert one $h'$ with v.e.v. $v\sin\beta'/\sqrt2$,
taken with $F'=2,\ \gamma_{F'}=1$ and $F=\gamma_Y+2\,$. \,With the same difference $F(h')-F(h_{\rm sm}) =2$ as for
 $F(h_2)-F(h_1^c) =2\gamma_A=2$ before,
$m_U$ is still given, independently of $\gamma_Y,\,\gamma_B$ and $\gamma_{L_i}$, by eq.\,(\ref{mu20},\ref{mu4},\ref{mu2h}),
\be
\label{muhh'}
m_U\simeq\frac{g"\cos\xi}{2}\, \sqrt{v^2\sin^2 2\beta'+F_\sigma^2 w^2}=\dis \frac{g"\cos\xi}{2}\,\frac{v\sin2\beta'}{r} ,
\ee
leading as before to
\be
\label{rsemi}
\frac{g"\cos\xi}{4}\,\simeq\, \frac{m_U}{2v}\ \frac{r}{\sin 2\beta'}\,\simeq \,2\times 10^{-6}\ m_U(\hbox{MeV})\ \frac{r}{\sin 2\beta'}\ ,
\ee
\vspace{-3mm}

\noindent
just changing $\beta$ into $\beta'$.

\vspace{2mm}

All this remains true if we further replace $h_{\rm sm}$ by $h_2$ and $h_1^c$ (taken for $\gamma_A=0$ with the same $Y\!=1$ and $F\!=$ $\gamma_Y$),
 with v.e.v.'s $\,v_2/\sqrt 2$ and $v_1/\sqrt 2$ such that
 \be
 \label{vev2}
 \hspace{-4mm}\ba{c}
 \!\left\{\
 \ba{ccccl}
 v_1\!&=&\!u\,\cos\beta\!&=&\!v\,\cos\beta\, \cos\beta',
 \vspace{2mm}\\
 v_2\!&=&\!u\,\sin\beta\!&=&\!v\,\sin\beta\, \cos\beta'
 \vspace{2mm}\\
 v'\!&=&&&\ \ \ v\,\sin\beta'.
  \ea \right.\!\!\!\!\!\!
 \ea
\ee
In both cases  $\eta=2\,v'^2/v^2=2\,\sin^2\beta'$,
the mixing angle $\xi$ being given by
\be
\label{mix}
\tan\xi\,\simeq
\, \frac{g"}{g_Z}\,(\gamma_Y+2\,\sin^2\beta')\,.
\ee
and the mass $m_U$ by (\ref{muhh'}), with $m_Z\simeq m_{Z_\circ}/\cos\xi$\, as in (\ref{sum2},\ref{mz0}).

\vspace{2mm}

In these situations for which  $\gamma_A=0$ (with $h_{\rm sm},h'$ and $\sigma$, or $h_1^c,h_2,h'$ and $\sigma$),
 the axial part in the $U$ current, isovector, originates from the mixing with the $Z$, independently of $\gamma_Y,\gamma_B$ and $\gamma_{L_i}$. The $U$ charge of quarks and leptons  is a linear combination of $Q,\,B,\,L_i$ and $T_{3L}$, 
the current being given by eq.\,(\ref{juold4-1}) with $\gamma_A=0$, i.e.
\be
\label{ju6}
\ba{ccl}
{\cal J}_U^\mu\,\simeq \ 
g"\cos\xi\  \left[\,\gamma_Y\cos^2\theta\,J_{\rm em}^\mu +\frac{1}{2}\,(\gamma_B J^\mu_B + \gamma_i\,J^\mu_{L_i}) \,  \right.
\vspace{3mm}\\
\hspace{21mm}  +\ \eta\, \left. (J^\mu_{3L}-\sin^2\theta J^\mu_{\rm em})\,\right]\,.
\vspace{-4.5mm}\\
\ea 
\ee

\vspace{3mm}

\noindent
$\gamma_Y\cos^2\theta$ and $\eta$ provide, respectively, effective measures of the pure dark photon and dark $Z$ 
contributions to the $U$ current.

\vspace{2mm}
 
To illustrate explicitly the calculation of $m_U$, adding the ${\cal M}^2_\circ$ matrix obtained
from $h_{\rm sm}$ alone, or $h_1^c$ and $h_2$, with $Y\!=1$ and $F=\gamma_Y$,
\be
{\cal M}_\circ^2=\ \frac{u^2}{4}\left(\ba{cc}
g_Z^2 &  -\, g_Z \,g"\,\gamma_Y
\vspace{2mm}\\
-\,g_Z\,g" \,\gamma_Y &   \ \ g"^2 \,\gamma_Y^2 
\ea\right) ,
 \ee 
 that would lead to a massless $U$,
 the similar one from $h'$ (with $Y\!=1$ and 
 $F=\gamma_Y + 2\,)$, 
 \be
{\cal M}'^2= \frac{v'^2}{4}\left(\!\ba{cc}
g_Z^2 &\   - \,g_Z \,g"\,(\gamma_Y\!+2)
\vspace{2mm}\\
-\,g_Z\,g" \,(\gamma_Y\!+2) &\    \ \ g"^2 \,(\gamma_Y\!+2)^2
\ea\!\right) ,
 \ee
and the singlet contribution, we get as in (\ref{m2H})
\be
\label{m25}
\hbox{\small $\dis
{\cal M}^2\!= 
\frac{v^2}{4}\!
\left(\!\!\ba{cc}
g_Z^2 & \ - g_Z g"(\gamma_Y \!+\!2\sin^2\!\beta')
\vspace{2mm}\\
-g_Zg" (\gamma_Y\!+\!2\sin^2\!\beta')  &\  g"^2\,[(\gamma_Y\!+\!2\sin^2\!\beta')^2\! +\!H]
\ea\!\!\right)\!,
$}
 \ee
with

\vspace{-5mm}
\be
H\,=   \,\sin^2 2\beta'+ F_\sigma^2\ \frac{w^2}{v^2} \ .
\ee
This provides $\xi$ given in the small \,mass limit 
\vspace{-.3mm}
by (\ref{mix}),
\linebreak
$m_U\!\simeq\! (g"\!\cos\xi \,v/2) \sqrt H$ by (\ref{muhh'}), and $ m_Z\simeq g_Zv/(2\cos\xi)$ \linebreak as in (\ref{mz4}).

\section{\hspace{-1.5mm}\boldmath \vspace{2mm}The $\,U$ charge, \,grand-unification
and an \hspace{.3mm}electro\hspace{.3mm}strong\hspace{0.3mm} symmetry} 
\label{sec:gut}

\vspace{-.6mm}

Further restrictions may be obtained in the framework of grand-unification \cite{gg}, if we demand the extra-$U(1)$ generator $F$ to commute with $SU(5)$ within a $SU(5)\times U(1)_F$ gauge group, so that
\be
F\,=\, \gamma_A \,F_A + \gamma_Y\,[\,Y-\hbox{\small$\dis\frac{5}{2}$}\ (B-L)\,]+
\gamma_{F'} F'+ \gamma_d \,F_d\,.
\ee
The vanishing of the vector coupling of the $U$ to up quarks, for  $\sin^2\theta=3/8$, 
is an effect of a  $SU(4)_{\rm es}$ electro\-strong symmetry group within $SU(5)\times U(1)_F$ \cite{sguts2}, relating the photon with the eight gluons. $(u_L,\bar u_L)$ transforms as an electro\-strong sextet so that the up quark  contributions to the $U$ current should then be axial (or vanish identically in the small $m_U$ limit), as for the $Z$ current itself \cite{epjc}.
\,Indeed for $\gamma_A\neq 0$  the antiquintuplet and quintuplet  v.e.v.'s $\,<\!h_1\!>$ and $<\!h_2\!>\, $  break spontaneously \cite{sguts2}
\be
\framebox[8.2cm]{\rule[-.25cm]{0cm}{1.05cm} $ \dis
SU(5)\times U(1)_F\!\!\! \stackrel{\ba{c}\hbox{$<\!h_1\!>,\,<\!h_2\!>$}\vspace{1mm}\\ \ea }{\hbox{\large$\longrightarrow$}}  \!\!\!SU(4)_{\rm \,electrostrong}\,,\!
$}
\ee
giving masses to the quartet  $(Y^{\mp1/3},\,W^\pm)$ and singlet $Z$ and $U$ bosons, with the $U$ remaining light for small $g"$, and the $X^{\pm 4/3}$ massless at this stage. 
$SU(4)_{\rm es}$, commuting with $U(1)_Z\times U(1)_U$, is  further broken to $SU(3)_{\rm QCD}$ $\times U(1)_{\rm QED}$ by an adjoint  \underline{\hbox{\bf 24}} or an equivalent mechanism, as within $N=1$ or \hbox{$N=2$} super\-symmetric (or higher-dimensional) grand-unified theories.

\vspace{2mm}

With a single quintuplet $h_{\rm sm}$ acquiring a v.e.v. (or two or even four as in supersymmetric theories but with the same gauge quantum numbers) one has with $\gamma_A=0$  the symmetry breaking pattern
 \be
SU(5)\times U(1)_F\ \ \stackrel{\ba{c}\hbox{$<\!h_{\rm sm}\!>$}\vspace{1mm}\\ \ea }{\hbox{\large$\longrightarrow$}}\ \ 
SU(4)_{\rm es}\times U(1)_U\,.
\ee

\noindent
The $U$ can then stay massless, or acquire a mass, possibly very small if $g"$ is very small, from the v.e.v. of the dark singlet $\sigma$, if present.

\vspace{2mm}
In all such cases the extra-$U(1)$ and $Z$ currents, and resulting $U$ current obtained from their combination, are invariant, above the grand-unification scale, under the $SU(4)_{\rm es}$ electrostrong symmetry group 
relating the photon with the eight gluons. It is spontaneously broken 
into $SU(3)\times U(1)_{\rm QED}$ at the grand-uni\-fication scale, 
\vspace{-.3mm}
with the $X^{\pm 4/3}$ and $Y^{\pm1/3}$ gauge bosons acquiring large masses, related  in the simplest case by \cite{sguts,sguts2}
\be
m_Y=\sqrt{m_W^2+m_X^2}\ .
\ee
This construction also provides a natural solution for the triplet-doublet splitting problem in grand-unified theories, 
with the doublet mass parameters automatically adjusting to zero so as to allow for the electroweak breaking. This mechanism can be transposed to (or originates
 from)  supersymmetric grand-unified theories in higher di\-mensions, for which $m_X\! \approx \pi \hbar/Lc$ \cite{nob}.
\vspace{2mm}

Independently of grand-unification, and of the possible values of $\,g"$ and $m_U$, supersymmetry can provide  the associations \cite{sguts,sguts2}
\be
\framebox[8.6cm]{\rule[-.25cm]{0cm}{.7cm} $ \dis
U\ \leftrightarrow \ \hbox{2 (or 4) uinos}\ \leftrightarrow\ \hbox{1 (or 5) spin-0 BEH bosons}\,,
$}
\ee
all with mass $m_U$ as long as supersymmetry is unbroken; this is also in agreement with gauge-BEH unification, allowing to view 
BEH bosons as extra spin-0 states for massive spin-1 gauge bosons.
We then have to take into account additional superpotential and supersymmetry-breaking terms, allowing for very small $g"$ and $m_U$ even with a very large extra-$U(1)$ $\xi " D"$ term. (There is also the further possibility, especially for very small $g"$ in $N\!=2$ supersymmetric GUTs, of viewing $\xi"$ as a dynamical quantity, very large in the first moments of the Universe.)
\,Some of these uinos (and associated spin-0 bosons) may remain lighter than the $U$ through a see-saw mechanism, in the presence of susy-breaking gaugino or higgsino mass terms.

\vspace{3mm}

$(Q_U)_V$ and $(Q_U)_A$ in (\ref{quv}), now restricted by the grand-unification and electrostrong symmetries, depend only on the three parameters $\gamma_Y,\,\gamma_A$ and $\eta$, instead of up to seven before. They read, with $\gamma_{L_i}=-\,\gamma_B=\frac{5}{2}\,\gamma_Y$,
\be
\label{qua5}
 \framebox [8.3cm]{\rule[-2cm]{0cm}{3.5cm} $ \dis
\ba{c}
\hbox{GUT}: \hspace{68mm}
\vspace{2mm}\\
\left\{\ 
\ba{ccl}
(Q_U)_V\!\!&=&\gamma_Y\,\hbox{\Large$\left[\right.$}\cos^2\theta \ Q\,  -\,\dis\frac{5}{4}\ (B\!-\!L)
\hbox{\Large$\left.\right]$}
\vspace{1mm}\\
&& \!\dis +\ \eta\ \hbox{\Large$\left[\right.$} (\,\frac{1}{2}-\,\sin^2\theta\,)\ Q  -\,\frac{1}{4}\, (B-L)\hbox{\Large$\left.\right]$}\, ,\!\!
\vspace{2.5mm}\\
(Q_U)_A\!\!&=&\dis  \frac{\gamma_A }{2}\,F_A- \,\frac{\eta}{2}\ T_{3A}\ ,
\ea \right.
\vspace{-5.5mm}\\
\ea
$}
\ee

\vspace{2mm}
\noindent
with $(T_{3L})_A=-\,T_{3A}/2$.
The axial couplings, already given in (\ref{gaeb20}) independently of $\sin^2\theta$, and of grand-unifica\-tion, are the same for down quarks and charged leptons, as also required by the $SU(4)_{\rm es}$ electrostrong symmetry.

\vspace{2mm}
This one becomes manifest for the vector couplings for $\sin^2\theta$ having the grand-unification value 3/8 \cite{gg}. We then have 
\be
(Q_U)_V=  \dis(5\gamma_Y+\eta)\ \,\frac{Q-2\,(B-L)}{8}\,=\,(5\gamma_Y+\eta) \ Q_{Z\,V}^{\rm gut}\,.
\ee


\noindent
It is proportional to the $SU(4)_{\rm es}$ invariant combination appearing in (\ref{qzgut}),
and as the weak charge $(Q_Z)_V=Q_{\rm Weak}/4$ in (\ref{qzu},\ref{qzu2},\ref{qzu8}) \cite{epjc},

\vspace{-4mm}
\be
\ba{c}
Q_{Z\,V}^{\rm gut}=\dis \frac{Q-2\,(B-L)}{8}=\,\left\{\ba{clc}
\ \ \ 0 &\hbox{for \underline{\boldmath $ 6$} = $(u,\bar u)_L$}\,,
\vspace{2mm}\\
-\,\frac{1}{8} &\hbox{for \underline{\boldmath $ 4$} =  $(d,e^+)_{L+R}$}\,,
\vspace{2mm}\\
+\,\frac{1}{4} &\hbox{for \underline{\boldmath $ 1$} =  \,$\nu_L\ [\,+\,\nu_R\,$]}\,.
\ea \right.
\vspace{1mm}\\
\ea
\ee

\vspace{0mm}
\noindent
It vanishes, as required for a vectorial charge, for the $(u,\bar u)_L$ sextet  (the vector coupling at the GUT scale then appearing as ``$u$-phobic'').

\vspace{2mm}

At the same time $(Q_U)_A$ in (\ref{qua5})
involves $F_A$ and $T_{3A}$, both invariant under the $SU(4)_{\rm es}$ electrostrong symmetry group, 
with

\vspace{-6mm}
\be
F_A\,=\,-\frac{1}{2} \ \ \hbox{for }\ \left\{\ba{clc}
 &\hbox{\underline{\boldmath $ 6$} = $(u,\bar u)_L$}\,,
\vspace{2mm}\\
 &\hbox{\underline{\boldmath $ 4$} =  $(d,e^+)_{L}$}\,,
\vspace{2mm}\\
 &\hbox{\underline{\boldmath $ \bar 4$} =  $(\bar d,e^-)_{L}$}\,,
\vspace{2mm}\\
 &\hbox{\underline{\boldmath $ 1$} =  \  \, $\nu_L\ [\,+\,\bar \nu_L\,]$}\,,
\ea \right.
\ee

\vspace{-1mm}

\noindent
and
\vspace{-6mm}

\be
T_{3A}=T_{3R}-T_{3L}\,=\,\left\{\ba{clc}
 -\,\frac{1}{2} &\hbox{for \underline{\boldmath $ 6$} = $(u,\bar u)_L$}\,,
\vspace{2mm}\\
+\,\frac{1}{2} &\hbox{for \underline{\boldmath $ 4$} =  $(d,e^+)_{L}$}\,,
\vspace{2mm}\\
+\,\frac{1}{2} &\hbox{for \underline{\boldmath $ \bar 4$} =  $(\bar d,e^-)_{L}$}\,,
\vspace{2mm}\\
-\,\frac{1}{2} &\hbox{for \underline{\boldmath $ 1$} = \  \,$\nu_L\ [\,+\,\bar \nu_L\,]$}\,.
\ea \right.
\ee
This shows how a new $U$ boson, originating from an extra $U(1)_F$ symmetry commuting with $SU(5)$
and mixed with the $Z$, is associated with a $U(1)_U$ symmetry  commuting with 
a $SU(4)_{\rm es}$ electrostrong symmetry group associating the photon with the eight gluons \cite{epjc,sguts2}.
For $\,\gamma_Y\!=\gamma_B=\gamma_{L_i}\!=0$  we still have a situation compatible with grand-unification with $Q_U$ depending only on $\gamma_A$ and $\eta$ as in \cite{plb86} (or just on $\eta$ as for a dark $Z$), its vectorial part becoming almost protophobic for $\sin^2\theta$ close to 1/4.

\vspace{-1.5mm}

\section{Conclusions}

\vspace{-1mm}

The standard model of particle physics is a great construction for understanding the world of particles and interactions. But it leaves many questions unanswered, and cannot be taken as a complete theory.
Among them,  the nature of dark matter, possibly made of new particles not present in the SM, and of dark energy. The latter  is related with the cosmological constant and vacuum energy density, extremely large in the very first moments of the Universe but extremely small now.
Furthermore, if we know four types of interactions,  other ones may well exist.

\vspace{1.5mm}

Essentially all extensions of the standard model motivated by these questions involve new particles and fields, and new symmetries.
The main approach has been, for several decades, to turn of higher energies to search for new heavy particles.
The discovery of the BEH boson at 125 GeV was a triumph for the standard model, but  superpartners in its supersymmetric extensions have not been found yet, and compactification scales $\approx \hbar/Lc$ may well be far away. So, if there must be ``new physics'', where is it ?
\vspace{1.5mm}

A complementary approach, which  has grown into a subject of  intense interest, involves moderate or even low energies, searching for new interactions and their mediators. This may involve interactions significantly weaker than weak interactions, down possibly to the strength of gravity and even less.

\vspace{1.5mm}
Using symmetries has good chances to remain a valid guiding principle for discussing new particles and interactions, all four known ones being associated with gauge or space-time symmetries. 
Supersymmetry provides an extension of space-time to fermionic coordinates, with a new $U(1)_R$ symmetry
acting chirally on them. It leads to $R$ parity, $R_p=(-1)^R$,  
at the origin of the stability of the lightest supersymmetric particle and of its possible role as a dark matter candidate.

\vspace{1.5mm}
Just as $U(1)_R$ acts chirally on gauginos and higgsinos, one can consider a $U(1)_A$ symmetry acting chirally on matter fields, as also suggested by the 2-BEH doublet structure of supersymmetric theories. Both chiral symmetries play an important role in discussions of spontaneous supersymmetry breaking, possibly  through the $\xi D$ term of an extra $U(1)$. And they might get related within extended supersymmetry, enlarging $U(1)_R$ up to $SU(4)$.

\vspace{1.5mm}

One may also consider extra space dimensions, which naturally combines with supersymmetry. The possibility of performing translations or moving along compact dimensions in the higher-$d$ space-time may result in additional 
$U(1)$'s, as associated with (central) charges in the extended supersymmetry algebra, and gauged by graviphotons that may be described by extra components of the gravitational field. Many such $U(1)$'s tend to be generated from (super)string theories, and it is natural to expect their remnants to act in 4 dimensions.

\vspace{1.5mm}

We may also extend directly the SM group beyond $SU(3)\times SU(2)\times U(1)_Y$, or $SU(5)$ for a grand-unified theory.
The minimal approach involves an extra-$U(1)_F$ symmetry, focussing here on small values of  its gauge coupling $g"$. 
Its possible generators naturally appear as linear combinations of $Y, \,B$ and $L$, or $B-L$, with an axial $F_A$.
$\ Y-\frac{5}{2}\,(B-L)$ and $F_A$ have a special status as they both commute with the $SU(5)$ grand-unification group.
The resulting $U(1)_U$ generator commutes with a $SU(4)_{\rm es}$ electrostrong symmetry group, spontaneously broken 
to $SU(3)\times U(1)_{\rm QED}$ at the grand-uni\-fication scale, possibly through extra dimensions.

\vspace{2mm}

These extra $U(1)$'s can provide a bridge to a new dark sector, allowing for dark matter particles to annihilate and opening the possibility for them to be light.
The associated light neutral gauge boson, called $U$ forty years ago, may 
manifest under different aspects, depending on its mass, the vector and/or axial character of its couplings, the strength of its interactions, its lifetime and decay modes, etc.

\vspace{2mm}

We have presented a unified framework relating the different aspects of such a new gauge boson.
Its possible couplings appear in the visible sector as linear combinations of $Q,\ B$ and $L_i$ with the weak isospin $T_{3L}$ and an axial generator $F_A$.
This relies on a minimal set of hypothesis, illustrating the crucial role of the BEH sector in determining the properties of the new boson.
Its current is a linear combination of the extra $U(1)_F$ and $Z$ currents, determined by
the $Z-U$ mixing angle $\xi$. This one,  small if the $U$ is light, is given in this limit by a sum on BEH doublets, with
$\,\tan\xi\,\simeq ({g"}/g_Z)\,\sum_i \epsilon_i \,F(h_i)\, {v_i^2}/{v^2}=({g"}/{g_Z)}$ $\, (\gamma_Y+\eta)\,,$
so that (disregarding for simplicity $\cos\xi\simeq 1$)
\be
\ba{c}
\ba{ccl}
\!{\cal J}^\mu_U\!&\simeq &g"\ \dis \hbox{\Large $\left(\right.$} \ \frac{1}{2}
\ J^\mu_F+(\gamma_Y+\eta)\,(J^\mu_{3L}-\sin^2\theta\,J^\mu_{\rm em})\,\hbox{\Large$\left.\right)$}
\vspace{2mm}\\
& \simeq & \dis g"\,\left(\,\gamma_Y J^\mu_{\rm em}+\eta\,J^\mu_Z +\frac{\gamma_B}{2}\,J^\mu_B+\!\frac{\gamma_{L_i}}{2}\,J^\mu_{L_i} \!+\!\frac{\gamma_A}{2}\,J^\mu_A\,\right) 
\vspace{2.5mm}\\
&& \ \ \ \ \ \ \ \ \ \ +\ \, {\cal J}^\mu_{\rm \,dark}\,+\,...\ .
\ea
\vspace{-3.5mm}\\
\ea
\ee

\vspace{3mm}

The $U$ boson may manifest as a generalized dark photon also coupled to $B$ and $L_i$, a dark $Z$ boson, 
an axial boson, and   a   (quasi-``invisible'') axionlike particle.
It connects these different aspects and interpolates between them, the dark photon case corresponding for example to the specific direction $(1,\,0,\,0,\!0,\!0\,;\,0,0)$ in a 7-dimensional space parametrized by
       \be
       \label{seven}
       \gamma_Y,\ \gamma_B,\, (\gamma_{L_e},\gamma_{L_\mu},\gamma_{L_\tau});\, \gamma_A\ 
       \hbox{and} \ \eta\,.
       \ee
The $U$ may also couple to dark matter (in particular if the $R$ current is involved in the gauging), possibly more strongly than to ordinary particles. Invisible decay modes into light dark matter may be then favored.

       \vspace{2mm}
       
       In the visible sector we have the correspondences:
       \be
       \framebox [8.55cm]{\rule[-1.8cm]{0cm}{3.9cm} $ \dis
       \ba{ccc}
       \gamma_Y &\to&\!\!\hspace{-5mm}\hbox{dark photon,}
       \vspace{3mm}\\
            \eta &\to&\hspace{-8mm}\hbox{dark}\ Z\ \
       \vspace{2mm}\\
        \gamma_A &\to&\hspace{-8mm}\ba{c}\hbox{isoscalar}\vspace{.5mm}\\ \hbox{axial boson}\ea \ \
       \vspace{3mm}\\
           \gamma_B =-\,\gamma_L\hspace{-10mm}&& \hspace{4mm} \to\ \  B\!-\!L\ \, \hbox{gauge boson,}\!\!\!\!\!\!
       \vspace{2mm}\\
      \hspace{-5mm}&& \hbox{\small etc.} \ .
       \ea 
       \hspace{-13mm}
       \ba{c}
      \vspace{-9mm}\\
       \left\}\ \to\!
       \ba{ccc}
        \hbox{(quasi-``invisible'')}
       \vspace{1mm}\\
       \hbox{axionlike behavior}
       \vspace{1mm}\\
       \hbox{\ of longitudinal} \ U
       \ea \right.
       \ea
       $}
       \ee
       \vspace{-3mm}

       \pagebreak
       
       \noindent
The $U$ charge is expressed for vector couplings  as a combination of $Q$ with $B$ and $L_i$, or $B-L$ in a grand-unified theory,
\be
\ba{ccl}
(Q_U)_V\!\!&=&\gamma_Y\cos^2\theta \ Q\,  +\,\dis\frac{\gamma_B}{2}\ B\,+\frac{\gamma_{L_i}}{2}\ L_i
\vspace{1mm}\\
&& \!\dis +\ \eta\ \hbox{\Large$\left[\right.$} (\,\frac{1}{2}-\,\sin^2\theta\,)\ Q  -\,\frac{1}{4}\, (B-L)\hbox{\Large$\left.\right]$}\, .
\ea
\ee
When the $\eta$ term from the mixing with $Z$ dominates, the $U$ has a nearly protophobic behavior, with
$(Q_U)_V$ almost proportional to $B-L-\,Q$, which may favor a small $\pi^0\to\gamma\,U$ decay rate.
In the framework of grand-unification, all vector and axial couplings of the $U$ may be expressed  as in (\ref{qua5}) in terms of three parameters only, $\gamma_Y, \gamma_A$ and $\eta$, instead of the seven ones in (\ref{seven}).

\vspace{2mm}

Axial couplings

\vspace{-5mm}
\be
\label{gapm}
g_A\simeq\, g" \,(Q_U)_A\simeq  \dis \, \frac{g"}{4}\,(\gamma_A\mp\eta)\,,
\ee
may be isoscalar,
or isovector from the contribution of the $Z$ current, or a mixing of both, as with
$
g_{A}\simeq ({g"}/4)\, (1\mp \cos 2\beta)
$
 in a two-doublet + one-singlet model.
The extra-$U(1)$ gauge coupling $g"$, then proportional to $m_U$ and to the invisibility parameter $r=\cos\theta_A$, is given by
 \be
\label{g"mu}
\frac{g"}{4}\,\simeq\,\frac{m_U}{2v}\ \frac{r}{\sin2\beta}\,\simeq \,2\times 10^{-6}\ \frac{r}{\sin2\beta}\ m_U(\hbox{MeV})\,.
\ee
 It is naturally very small if the $U$ is very light, or if $r$ is small from a large singlet v.e.v., with the extra-$U(1)$ symmetry broken at a large scale.
 
\vspace{2mm}

 Axial couplings of the $U$, when present, are responsible for  enhanced interactions of a longitudinal $U$, 
interacting in the small $m_U$ limit with
 effective pseudoscalar couplings 
 \vspace*{-2mm}
 \be
 \ba{ccl} \hspace{-4mm}
 \dis g_P\!&=& \dis g_A\ \frac{2m_{ql}}{m_U}\,=\, 2^{1/4}\,G_F^{1/2}\,A_\pm 
 \vspace{0mm}\\
 &\simeq& 4\times 10^{-6}\,m_{ql}\hbox{(MeV)} \, \times
\,  \left\{\!\ba{c} r\,\cot\beta\,,\vspace{2mm}\\ r\,\tan\beta\,. \ea\right.\!\!
  \ea
 \ee
 \vspace{-3mm}
 
 \noindent
This  reconstructs the same pseudoscalar couplings as for a quasi-invisible axion, with $r=\cos\theta_A$ as the invisibility parameter. With a semi-inert doublet $h'$, one has, similarly,
$
\,g_P
 \simeq \mp \ 4\times 10^{-6}\,m_{ql}\hbox{(MeV)} \,r\,\tan\beta'.
$

\vspace*{2mm}

$\psi,\,\Upsilon,\,K$ and $B$ decays lead to strong limits on axial couplings, typically $\simle 2\times 10^{-7}\,m_U$ (MeV) or even down to $\simle 2\times 10^{-9}\,m_U$ (MeV). Such a strong limit on $g_{Ae}$ may facilitate satisfying the atomic-physics  parity-vio\-lation constraint.
The $U$ couplings to neutrinos may also be very small, thanks to a small $g"$ naturally proportional to $m_U$ and $r$; 
 or to a small 
$Q_U(\nu_L)$, if $Q_U$ is close to a combination of $Q,\,B,\,T_{3R}$ and $L+2F_A$.
The neutrino and electron couplings, proportional to $m_U\,r\, Q_U$, are constrained from $\nu$-$e$ scattering experiments to
$\sqrt{\,g_{\nu_L}g_e} \,\simle \,(1\ {\rm to}\ 3) \times 10^{-6}\ m_U(\hbox{MeV})$
for $m_U \simge $ a few MeV's, much in line with expression (\ref{g"mu}) of $g"/4$ as $2\times 10^{-6}\ m_U(\hbox{MeV})
 \times {r}/{\sin2\beta}$\,.

\vspace{2mm}

The extra-$U(1)$ coupling $g"$  may vary between  sizeable values for larger $m_U$'s $\sim $ a few ten GeV's, down to $\simle 10^{-24}$ for an extremely light or even massless $U$ boson. It could then be detectable through apparent violations of the Equivalence principle. The extreme weakness of such a new long range force might be related, within supersymmetry, with a very large energy scale $\simge 10^{16}$ GeV, associated with a huge vacuum
energy density that may be at the origin of the inflation of the early Universe.

\vspace{3mm}

The possible existence of a new interaction, that may fit within the grand-unification framework and have connections with the fundamental structure of space-time, is both a fascinating theoretical subject and a very rich field for experimental investigations.

\end{document}